\documentclass[10pt,aps,prd,twocolumn,preprintnumbers,superscriptaddress,amsmath,amssymb,nofootinbib,floatfix]{revtex4-1}

\usepackage[hidelinks]{hyperref}
\usepackage{mathtools}
\usepackage{aas_macros}
\usepackage[capitalize]{cleveref}
\usepackage{amsmath}
\usepackage{amssymb}
\usepackage{amstext}
\usepackage{graphicx}
\usepackage{siunitx}

\Crefname{equation}{Equation}{Equations}
\crefname{equation}{Eq.}{Eqs.}
\Crefname{figure}{Figure}{Figures}
\crefname{figure}{Fig.}{Figs.}
\Crefname{table}{Table}{Tables}
\crefname{table}{Tab.}{Tabs.}
\Crefname{section}{Section}{Sections}
\crefname{section}{Sec.}{Secs.}

\newcommand{\p}{\mathrm{p}}

\newcommand{\prm}[1]{{#1}^{\prime}}
\newcommand{\dprm}[1]{{#1}^{\prime\prime}}
\newcommand{\sft}{\beta}
\renewcommand{\d}[2][]{\operatorname{d}^{#1}\!{#2}}

\newcommand{\conformalH}{\mathcal{H}}
\newcommand{\deriv}{\prm}
\newcommand{\dderiv}{\dprm}

\usepackage{xcolor}

\hypersetup{colorlinks, linkcolor={blue}, citecolor={blue}, urlcolor={blue}}
\usepackage[font=small, labelfont=bf, justification = raggedright]{caption}
\usepackage{graphicx}

\let\subfloat\relax
\usepackage{subcaption}
\usepackage{bm}
\usepackage[bottom]{footmisc}

\begin{document}

\preprint{Prepared for submission to Phys. Rev. D}
\title{Analytical approximations for curved primordial tensor spectra}

\author{Ezra Msolla}
\email[]{ezra.msolla@alumni.utoronto.ca}
\affiliation{David A. Dunlap Department of Astronomy and Physics, University of Toronto, 50 St.George Street, Toronto, ON M5S 3H4, Canada}
\affiliation{Texas Center for Cosmology and Astroparticle Physics, Weinberg Institute,Department of Physics, University of Texas at Austin, Austin, TX 78712, USA}

\author{Ayngaran Thavanesan}
\email[]{at735@cantab.ac.uk}
\affiliation{Department of Applied Mathematics and Theoretical Physics, University of Cambridge, Wilberforce Road, Cambridge, CB3 0WA, UK.}
\affiliation{Kavli Institute for Theoretical Physics, Santa Barbara, CA 93106, USA.}
\affiliation{Laboratory for Theoretical Fundamental Physics, Institute of Physics, École Polytechnique Fédérale de Lausanne (EPFL), CH-1015 Lausanne, Switzerland.}

\date{\today}

\begin{abstract}
    \vspace{5pt}
    We build upon previous analytical treatments of scalar perturbations in curved inflationary universes to obtain analytical templates for the primordial tensor power spectrum in models with non-zero primordial spatial curvature. These templates are derived without assuming a particular inflaton potential, within a background history consisting of an initial kinetically dominated phase followed by an ultra-slow-roll inflationary regime, thereby isolating the universal imprints of curvature on tensor modes. Our results predict characteristic large-scale features --- including low-$\ell$ cut-offs and oscillatory patterns --- that are consistent with numerical solutions and provide a clear physical interpretation of how curvature modifies the underlying dynamics. In particular, we show that curvature effects manifest mathematically as systematic shifts in the dynamically relevant wavevectors, mirroring the behaviour previously identified in the scalar power spectrum. These features translate into distinctive signatures in the large-angle $B$-mode polarisation spectrum, offering a potential discriminant for spatial curvature in forthcoming CMB observations.
    \vspace{20pt}
\end{abstract}
\maketitle

\section{Introduction}\label{sec:introduction}
The inflationary paradigm remains the most successful framework for explaining the observed flatness, homogeneity, and isotropy of the universe, while naturally generating the primordial perturbations that seeded large-scale structure~\cite{Guth:1980zm,Linde:1981mu,Starobinsky:1980te}. These perturbations produced the temperature anisotropies measured in the cosmic microwave background (CMB), upon which the $\Lambda$CDM model is based, describing the fluctuations as nearly scale-invariant, adiabatic, and Gaussian.

If an inflationary phase is introduced to explain current observations, one cannot a priori assume that the universe was spatially flat at its onset. Inflation generically suppresses, but does not entirely erase, curvature since it must eventually end. Consequently, small but non-negligible residual curvature may survive, consistent with observational limits. For levels of curvature permitted by present CMB bounds, the resulting scalar and tensor spectra typically display large-scale power suppression and oscillatory features. Bayesian analyses have explored the degree of fine-tuning required in such scenarios~\cite{Hergt:2018ksk,Hergt:2022fxk}, while numerical studies demonstrate that curved-inflationary models can naturally generate these features~\cite{Agocs:2019vyk,Handley:2019}, potentially alleviating tensions between early- and late-universe datasets~\cite{Riess,Joudaki:2016mvz,Kohlinger:2017sxk,Hildebrandt:2016iqg,DES:2017myr,Handley:2019wlz,Philcox:2020vvt,Ivanov:2020mfr}.

Models with spatial curvature have also been revisited in the context of late-time cosmological tensions. Extensions of $\Lambda$CDM that allow for non-zero present-day curvature --- the so-called $K\Lambda$CDM cosmologies~\cite{Ellis:2002we,Lasenby:2003ur} --- can yield improved fits to present data (see e.g.~\cite{Handley:2019,Specogna:2025ufe}). The detection of even small curvature today would place strong constraints on inflationary dynamics, motivating ``just-enough'' inflation models~\cite{Handley:2014bqa,Hergt:2018crm,Blachier:2023ooc,Schwarz:2009sj,Cicoli:2014bja,Boyanovsky:2006qi,Boyanovsky:2006pm}.

While scalar perturbations in curved inflationary universes have been analytically modelled~\cite{Ratra:1994vw,Bucher:1994gb,White:1994,Ratra:1995,Linde:1995,Bucher:1995,Ellis:2002we,Lasenby:2003ur,Efstathiou:2003tv,Lasenby:2005ig,Handley:2019,Avis:2019eav,Agocs:2020yjm,Shumaylov:2021qje,Letey:2022hdp,Specogna:2025ufe} in a potential-independent manner~\cite{Thavanesan:2020lov}, corresponding tensor mode analyses have been less developed~\cite{Bonga_2016,Bonga_2017,Franco:2017pxt,DAgostino:2023tgm}. In this paper, we extend the analytic framework of~\cite{Contaldi:2003zv,Thavanesan:2020lov} to derive compact approximations for the primordial \emph{tensor} power spectrum in the presence of spatial curvature. Our approach adopts that of \citet{Thavanesan:2020lov} and models the background as initially kinetically dominated before a sharp transition to an effectively potential-independent (``ultra-slow-roll'') phase generalising previous work for flat inflating universes~\cite{Contaldi:2003zv,Gessey-Jones:2021yky,Dineen:2023nbt}.
The adoption of this two phase inflationary is therefore a controlled idealisation that renders the problem analytically traceable and allows curvature effects to be isolated in a model independent manner.
Despite this idealisation, the analytic templates qualitatively reproduce the spectra obtained from full numerical computation while providing physical intuition for how curvature modifies the propagation of tensor modes.

Recent analyses of cosmological datasets continue to leave open the question of primordial curvature, with several studies identifying the tensor spectrum as a key observable that could help resolve this ambiguity~\cite{Specogna:2025ufe}. However, only a handful of precise calculations of the tensor spectrum in curved universes have been carried out, limiting our ability to assess how spatial curvature influences the amplitude and scale dependence of primordial gravitational waves. Extending the potential-independent treatment to tensor perturbations, therefore represents a natural and timely next step, which we undertake in this work.

Section \ref{sec:background} reviews the background equations in conformal time and derives the tensor mode equation of motion in curved inflationary spacetimes. In Section \ref{sec:computingCurvedSpectra}, we present the analytic solutions, compute the corresponding spectra, and compare them to numerical results. Section \ref{sec:discussion} concludes with a discussion of the implications for inflationary model building and observational constraints. Supplementary material, such as PYTHON code for generating figures and Mathematica scripts for computer algebra, is found at~\cite{msolla_2025_17593617}.

\section{Background}\label{sec:background}
In this section for completeness we derive the first-order tensor perturbation equations in conformal time from the Einstein equations, including a full derivation of the background Friedmann equations, using the approach of~\cite{Handley:2019,Thavanesan:2020lov}. 

The action of a minimally coupled scalar to a curved background in $D=3+1$ spacetime dimensions is
\begin{equation}\label{eqn:ScalarFieldAction}
    S=\int d^4\mathbf{x} \sqrt{-g} \Biggl\{\frac{M_{p}^{2}}{2}R +\frac{1}{2}g^{\mu\nu}\nabla_{\mu}\phi\nabla_{\nu}\phi - V(\phi) \Biggr\} \, .
\end{equation}
Extremising this action yields the Einstein field equations and a conserved stress energy tensor. 
\begin{equation}
 \label{eqn:EFEs}
    \begin{split}
          M_{p}^{2}G_{\mu\nu} &= T_{\mu\nu} \, , \\
     T_{\mu\nu} & = \nabla_{\mu}\phi\nabla_{\nu}\phi - g_{\mu\nu}\left( \frac{1}{2}\nabla_{\alpha}\nabla^{\alpha}\phi - V(\phi)\right) \, , \\
    \end{split}
\end{equation}
where $M_{p}$ is the reduced Planck mass. Throughout this paper, consistent with the cosmological principle, we shall assume that to zeroth order the solutions to these equations are homogeneous and isotropic. Concretely, one finds the stress-energy tensor has the standard form expected for a scalar field. We then perturbatively expand these equations about the homogeneous solutions to first order in the Newtonian gauge.

In spherical polar coordinates the conformal time metric is therefore
\begin{equation}
 \label{eqn:curvedmetric}
    \begin{split}
          ds^{2} = a(\eta)^{2}[(1 + 2\Phi)d\eta^{2} - (1-2\Psi)(c_{ij} + h_{ij})dx^{i}dx^{j}], \\
      c_{ij}dx^{i}dx^{j} = \frac{dr^{2}}{1-Kr^{2}} + r^{2}(d\theta^{2} +\sin^{2}\theta d\phi^{2}),
    \end{split}
\end{equation}
where $a(\eta)$ is the usual scale factor in cosmology, $K$ denotes the sign of the spatial curvature, taking values of $K = +1$ for a closed (positively curved) universe, $K = -1$ for an open (negatively curved) universe, and $K = 0$ for a flat universe. The potentials $\Phi$ and $\Psi$ along with $\delta\phi$ are scalar perturbations, whilst $h_{ij}$ is a divergenceless, traceless tensor perturbation with two independent polarisation degrees of freedom.

\subsection{Zeroth order equations}
At zeroth order in the perturbations, the (00)-component of the Einstein field equations and the (0)-component of the conservation of the stress-energy tensor \eqref{eqn:EFEs} give the evolution equations of the homogeneous Friedmann-Lemaître-Robertson-Walker (FLRW) spacetime with matter content defined by a scalar field. 
Due to the homogeneity of the scalar field, its spatial derivatives vanish. Computing the (00)-stress-tensor yields
\begin{equation}
 \label{eqn:T00component}
    T_{00} =\tfrac{1}{2}(\phi')^2 + a^2 V(\phi) \, .
\end{equation}
Furthermore, we explicitly write the (00)-Einstein tensor as
\begin{equation}
 \label{eqn:G00component}
    G_{00} = 3\left( \mathcal{H}^{2} +K \right) \, .
\end{equation}
Inserting Eqs. \eqref{eqn:T00component} and \eqref{eqn:G00component} into \eqref{eqn:EFEs} and raising the first index, we are able to derive the first of the background equations
\begin{equation}
 \label{eqn:Friedmann1}
     \mathcal{H}^{2} +K = \frac{1}{3M_{p}^{2}}\left(\frac{1}{2}(\phi')^2 + a^{2}V(\phi)\right) \, ,
\end{equation}
where $\mathcal{H}= a'/a$ is the conformal Hubble parameter, $V(\phi)$ is the scalar potential, and $a$ is the scale factor and primes indicate derivatives with respect to conformal time $\eta$ defined by $d\eta = dt/a$.

In terms of the scalar field and potential, the zeroth component of the conservation of the stress-energy tensor is
\begin{equation}\label{eqn:KleinGordon1}
    \left( \Box\phi - \frac{d}{d\phi}V(\phi) \right)\,\phi' =0 \, ,
\end{equation}
where $\Box= \nabla^{\mu}\nabla_{\mu}$ is the covariant d'Alambertian. Expanding and taking the conformal derivative of the scalar field to be nonzero implies the dynamical Klein–Gordon equation in curved spacetime
\begin{equation}\label{eqn:KleinGordon2}
    \phi'' + 2\mathcal{H}\phi' + a^{2}\frac{d}{d\phi}V(\phi) = 0 \, .
\end{equation}

A further useful relation is found by taking the derivative of \eqref{eqn:Friedmann1}
\begin{equation}
 \label{eqn:FriedmannDeriv}
    2\mathcal{H}\mathcal{H}'  =  \frac{1}{3M_{p}^{2}}\left(\phi'(\phi'' + a^{2}\frac{d}{d\phi}V(\phi)) + 2a^{2}\mathcal{H}V(\phi) \right) \, .
\end{equation}
One can identify and substitute the Klein-Gordon equation \eqref{eqn:KleinGordon2} into this expression and rearrange terms to derive the evolution equation for the conformal Hubble parameter
\begin{equation}
 \label{eqn:HubbleDerivative}
    \mathcal{H}' = -\frac{1}{3M_{p}^2}\left(\phi^{\prime 2} - a^2 V(\phi)\right) \, .
\end{equation}
Moving forward, Eqs. \eqref{eqn:Friedmann1} and \eqref{eqn:KleinGordon2} are used to remove all explicit potential dependency from the tensor equations of motion, and Eq. \eqref{eqn:HubbleDerivative} to remove all derivatives of $\mathcal{H}$ in place of $\phi$. 

\subsection{First order equations}
Considering Eq. \eqref{eqn:curvedmetric}, we initially work in proper time for ease of calculation, we immediately see the spatial part of the metric is 
\begin{equation}
 \label{eqn:spatial metric}
    g_{ij} = -a^{2}(c_{ij} + h_{ij}) \, ,
\end{equation}
and the first order perturbative Einstein Equations given by
\begin{equation}
    \label{eqn:PerturbEFEs}
     \delta G^{i}_{\; j} = \delta R^{i}_{\; j} \, ,
\end{equation}
where $R_{ij}$ is the Ricci tensor and were considering variations w.r.t the spatial metric.

In order to derive the equations of motion for the tensor perturbations, we first derive the non-zero Christoffels, which enter the first order Einstein equations as
\begin{equation}
 \label{eqn:Christoffels}
    \begin{split}
    \Gamma^{0}_{\;ij} & =  -Hg_{ij} + \frac{1}{2}a^{2}\dot{h}_{ij} \, ,\\
    \Gamma^{i}_{\;0j} &=  H\delta^{i}_{\ j} + \frac{1}{2}\dot{h}^{i}_{\ j} \, ,\\  
    \Gamma^{i}_{\;jk} &= \frac{1}{2}(\nabla_{k}h^{i}_{\ j} + \nabla_{j}h^{i}_{\ k} - \nabla^{i}h_{jk}) \, .
\end{split}
\end{equation}
Additionally, the Spatial components of Ricci read as
\begin{equation}
 \label{eqn:Ricci}
    \begin{split}
        R_{ij} &= -\left( \frac{\ddot{a}}{a}+ 2H^{2} + \frac{2K}{a^2}\right)g_{ij} +  \frac{a^{2}}{2}\bigg[\ddot{h}_{ij} +3H\dot{h}_{ij}  \\
     & \hspace{4cm}-\left(\frac{1}{a^2}\nabla^{k}\nabla_{k} -\frac{2K}{a^2}\right)h_{ij}\bigg] .
    \end{split}
\end{equation}
Substituting this expression into Eq. \eqref{eqn:PerturbEFEs}, and enforcing the traceless-transverse gauge condition with no anisotropic stress which requires that the left hand side must vanish, yields the equation of motion for tensor perturbations
\begin{equation}
 \label{eqn:Tensormode}
      \ddot{h}_{ij} + 3H\dot{h}_{ij} - \frac{1}{a^{2}}(\nabla^{k}\nabla_{k} -2K)h_{ij} = 0 \, .
\end{equation}
Our goal now is to write this equation in conformal time and in a more familiar form. Explicit calculations of the first two terms which encode the time evolution of the tensor perturbations simplify to
\begin{equation}
    \label{eqn:conformalmode}
    \begin{split}
     \ddot{h}_{ij} &=   -\frac{a'}{a^{3}}h_{ij}' + \frac{1}{a^{2}}h_{ij}'' \, ,\\
    3H\dot{h}_{ij} &= 3\frac{a'}{a^{3}}h_{ij}' \, .
    \end{split}
\end{equation}
A conformal transformation, as well as a Fourier decomposition acts to replace the spatial covariant derivative with its associated scalar wavevector expression.
\begin{equation}
 \label{eqn:conformalwavevector}
    \begin{split}
        & \quad h_{ij}'' + 2\mathcal{H}h_{ij}' + (\mathcal{K}^{2} + 2K)h_{ij}=0 \, , \;\;\;\;\; \\
        &  \mathcal{K}^{2}(k) = \begin{cases}
k^2, & k \in \mathbb{R},\; k > 0,\; K = 0, -1 \, , \\[6pt]
k(k+2), & k \in \mathbb{Z},\; k > 2,\; K = +1~ \, .
\end{cases}
    \end{split}
\end{equation}
The tensor perturbations can be quantised in a similar fashion as the curvature perturbations in \cite{Thavanesan:2020lov}, and the corresponding spectrum can also be defined in the same way. Writing $h = u/a$, then, in Fourier space, Eq. \eqref{eqn:conformalwavevector} describing the tensor perturbations reduces to
\begin{equation}
    \label{eqn:tensorMSE}
     u_{k}'' + \left(\mathcal{K}^{2} + 2K  -\frac{a''}{a} \right)u_{k} = 0 \, .
\end{equation}
This equation is analogous to the Mukhanov-Sasaki equation \cite{MUKHANOV1992203} for curvature perturbations but with a re-definition of variables. This computation has historically been performed for curvature perturbations by \cite{Zhang:2003eh,Gratton:2001gw,PhysRevD.96.103534,Bonga_2016,Bonga_2017,Akama:2018cqv,Ooba:2017ukj}.

\section{Analytical primordial tensor power spectra for curved universes}\label{sec:computingCurvedSpectra}
We compute primordial tensor power spectra for curved inflationary universes by extending the approximate analytic framework of \citet{Thavanesan:2020lov} to the tensor sector, constructing potential-independent templates for curved inflationary dynamics. The background evolution is modeled as a two-phase history: an initial, pre-inflationary kinetically dominated (KD) stage,
\begin{equation}
    \textbf{Kinetic Dominance (KD): } \quad \phi'^{2} \gg a^{2}V(\phi) \, ,
\end{equation}
followed by an instantaneous transition to a phase in which the scalar field motion has significantly slowed,
\begin{align}
    \textbf{Ultra-Slow-Roll (USR): } \quad &\mathcal{E}=\frac{\phi'^2}{2\mathcal{H}^2}\to0 \, , \\
    &\phi'^{2} \ll a^{2}V(\phi) \, ,
\end{align}
where $\mathcal{E}$ is the standard slow-roll parameter, and the standard slow-roll constraints to solve the horizon problem are satisfied. The USR phase therefore represents the onset of inflation in our analytic framework, where the expansion becomes accelerated ($\ddot{a} >0$). This two-phase approximation captures the essential features of curvature-induced modifications to the power spectra while remaining agnostic to the precise form of the potential $V(\phi)$. The framework can later accommodate potential dependence through higher-order corrections in the analytic solutions for curved inflationary dynamics, but we leave this for future work.

We first derive analytic solutions and power-series expansions for the background variables in both regimes in Section~\ref{sec:backdynamics}. In Sections~\ref{sec:kindominance} and~\ref{sec:uslowroll}, we obtain closed-form solutions to the tensor mode equations in each regime, matching them continuously at the transition point. The corresponding primordial tensor power spectra are then computed from the freeze-out amplitudes of the ultra-slow-roll phase in Section~\ref{sec:powerspectrumderivation}.

Throughout this work, we adopt the convention that conformal time $\eta=0$ corresponds to the initial singularity, such that $a(\eta=0)=0$. The scale factor $a(\eta)$ carries dimensions of length, meaning that $a \neq R/R_{0}$ and is not normalised to unity at the present day---that is, at redshift $z=0$, $a(z=0)\equiv a_{0}\neq1$. As emphasised by~\citet{Agocs:2019vyk}, this redefinition implies that the comoving wavenumber and the physical scale of the curvature perturbation today differ by a factor of $a_{0}$. This choice of normalisation proves convenient when comparing analytic predictions to numerical results for curved cosmologies, since it allows curvature effects to be tracked explicitly through the dependence on $a_{0}$ and $\eta$.

\subsection{Background dynamics}\label{sec:backdynamics}
The expressions \eqref{eqn:Friedmann1} and \eqref{eqn:HubbleDerivative} can be rearranged to forms from which we solve for the background variables $a$, $\mathcal{H}$ and $\phi$. By simply adding and subtracting the two, one arrives at the following forms
\begin{align} 
  \label{eqn:PotentialHubble}
     \mathcal{H}' + 2\mathcal{H}^2 + 2K &=   a^2 V(\phi) \, , \\
  \label{eqn:FieldHubble}
     \mathcal{H}' -  \mathcal{H}^2 - K &=  \frac{1}{2}\phi^{\prime 2} \, .
\end{align}

In the initial stages of kinetic dominance $\phi'^{2} \gg a^{2}V(\phi)$, the right hand side of Eq. \eqref{eqn:PotentialHubble} is negligible. Noting that $\mathcal{H} =a'/a$, $\mathcal{H}' = a''/a - \mathcal{H}^{2}$, and introducing the substitution $\mu(\eta) =a^{2}(\eta)$, we arrive at the following simplified differential equation
\begin{equation} 
 \label{eqn:KDscalefactor}
    \mu'' + 4K\mu = 0 \, .
\end{equation}
Similarly in the ultra-slow-roll regime, $\phi'^{2} \ll a^{2}V(\phi)$, we can neglect the right hand side of  Eq. \eqref{eqn:FieldHubble}. Defining now $\mathcal{H} = -\frac{\nu'}{\nu}$ so that $ \mathcal{H}' = -\nu''/\nu + \mathcal{H}^{2}$ we arrive at a second differential equation
\begin{equation}
 \label{eqn:USRscalefactor}
    \nu'' +K\nu = 0 \, .
\end{equation}
The solutions to both Eqs. \eqref{eqn:KDscalefactor} and \eqref{eqn:USRscalefactor} are dependent on the curvature of the cosmology. To capture this dependence we define
\begin{equation}
 \label{eqn:CurvatureParametercases}
    S_{K}(x) =
    \begin{cases}
        \sin(x) & K =+1 \, , \\[4pt]
        x & K=0 \, , \\[4pt]
        \sinh(x)& K=-1 \, .
    \end{cases}
\end{equation}
then solving Eq. \eqref{eqn:KDscalefactor} and expressing in terms of the scale factor yields $a \sim \sqrt{S_{K}(2\eta)}$ for the kinetically dominated phase. Likewise, solving Eq. \eqref{eqn:USRscalefactor} yields $a \sim 1/S_{K}(\eta)$ in ultra-slow-roll. Each solution carries two integration constants: one corresponding to an additive coordinate shift in $\eta$ and the other a linear scaling of $a$. Imposing continuity of $a$ and $a'$ at some transition time $\eta=\eta_{t}$ gives
\begin{equation}
 \label{eqn:SFcases}
    a =
    \begin{cases}
        \sqrt{S_{K}(2\eta)} & : 0\leq\eta < \eta_{t} \, , \\[4pt]
        [S_{K}(2\eta_t)]^{3/2}/S_{K}(3\eta_t -\eta) & :\eta_{t} \leq \eta < 3\eta_{t} \, ,
    \end{cases}
\end{equation}
with conformal coordinate freeze out into the inflationary phase occurring as $\eta \to 3\eta_{t}$. For the closed case ($K = +1$), the transition time is bounded by $\eta_{t} = \pi/4$. Beyond this limit, the Universe begins to collapse prior to the transition, signaling a breakdown of our approximation. The remaining background quantities may also be obtained in the kinematically dominated regime \cite{Handley:2014bqa},but we require only power series expansions, which read as
\begin{align}
 \label{eqn:LogolinearN}
    N &= N_{p} + \frac{1}{2}\log\eta - \frac{K}{3}\eta^{2} - \frac{2K^{2}}{45}\eta^{4}+\mathcal{O}(\eta^{6}) \, , \\
    \mathcal{H} &= N' = \frac{1}{2\eta} -\frac{2K}{3}\eta - \frac{8K^{2}}{45}\eta^{3} + \mathcal{O}(\eta^{5}) \, , \label{eqn:LogolinearHubble} \\
    a &= e^{N} = e^{N_{p}}\eta^{1/2} - \frac{e^{N_{p}}K}{3}\eta^{5/2} + \frac{e^{N_p}K^2}{90}\eta^{9/2} + \mathcal{O}(\eta^{11/2}) \, , \label{eqn:LogolinearSF}
\end{align}
where $N = \log a$ and $N_{p}$ a constant of integration. The complete derivation requires logolinear power series expansions\cite{Lasenby:2003ur,Handley:2019bzs}, detailed fully in Appendix \ref{sec:Appendix}.
\par

The ultra-slow-roll regime is characterised by $\phi'^{2} \ll a^{2}V(\phi)$; in our analysis we adopt the analytic form of the scale factor given in Eq. \eqref{eqn:SFcases}.

\subsection{Kinetic dominance}\label{sec:kindominance}
The tensor variable $u_{k}$ obeys Eq. \eqref{eqn:tensorMSE}. Combining Eqs. \eqref{eqn:LogolinearN} and \eqref{eqn:FieldHubble}, one finds that for the kinetically dominated regime
\begin{equation}
    \label{eqn:KDLogolinearexp} 
    2K  -\frac{a''}{a} = \frac{1}{4\eta^{2}} + \frac{10K}{3} + \frac{4K^2}{15}\eta^2 +\mathcal{O}(\eta^{4}) \, .
\end{equation}
 Substituting Eq. \eqref{eqn:KDLogolinearexp} into \eqref{eqn:tensorMSE}, the tensor mode equation in the kinetically dominated regime to first order in $K$, becomes
\begin{equation}
 \label{eqn:KDMSequation}
\begin{split}
u_{k}'' + &\left( k_{-}^{2} + \frac{1}{4\eta^{2}} \right) u_{k} = 0 \, , \\
k_{-}^{2} &= \mathcal{K}^{2}(k) + \frac{10K}{3}  + \mathcal{O}(K^2)\, .
\end{split}
\end{equation}
We find that the first-order effects of curvature on the tensor mode equation of motion Eq. \eqref{eqn:tensorMSE} in the kinetically dominated regime manifest themselves as an effective shift in the wavevector, dynamically acting on the mode frequency.

Furthermore, solving Eq. \eqref{eqn:KDMSequation} leads to the following expression for the tensor mode variable $u_{k}$ in the kinetically dominated regime
\begin{equation}
 \label{eqn:KDtensormode}
     u_{k} = \sqrt{\frac{\pi}{4}}\sqrt{\eta}\left[ A_{k}H_{0}^{(1)}(k_{-}\eta) + B_{k}H_{0}^{(2)}(k_{-}\eta)\right] \, ,
\end{equation}
where $H_{0}^{(1,2)}$ are zeroth-order Hankel functions of the first and second kind. Canonical normalisation imposes $|B_{k}|^{2} -|A_{k}|^{2}=1$ \cite{Handley:2016ods}. Following~\cite{Thavanesan:2020lov}, and \cite{Contaldi:2003zv,Sahni:1990tx} we choose initial conditions that select the right-handed mode, such that
\begin{equation}
 \label{eqn:KDinitialconditions}
    A_{k} = 0 \, , \qquad B_{k}=1 \, ,
\end{equation}
leaving open the possibility for future work to consider alterative initial conditions.

\subsection{Ultra-slow-roll}\label{sec:uslowroll}
In the ultra-slow-roll regime ($\eta \geq \eta_{t}$) we consider the alternate form of the scale factor $a$ in Eq. \eqref{eqn:SFcases}. Then to first order in curvature the relevant terms in the tensor mode equation of motion Eq. \eqref{eqn:tensorMSE} are of the form
\begin{equation}
 \label{eqn:USRScalefactorexp}
      \frac{a''}{a} = \frac{2}{(3\eta_{t}- \eta)^{2}} - \frac{K}{3} + \mathcal{O}((3\eta_{t}- \eta)^{2}) \, .
\end{equation}
Substituting this result into \eqref{eqn:tensorMSE} we can then express the tensor mode equation for the ultra-slow-roll regime as 
\begin{equation}
 \label{eqn:USRMSequation}
\begin{split}
  u_{K}'' + &\left[ k^{2}_{+} - \frac{2}{(3\eta_{t} -\eta)^{2}} \right]u_{K} = 0 \, ,\\
k^{2}_{+}& = \mathcal{K}^{2}(k) + \frac{7K}{3} + \mathcal{O}(K^2) \, .
\end{split}
\end{equation}
In addition to the shifted wavevector $k_{-}$ defined for the kinetic dominance regime ($\eta \leq \eta_{t}$), we introduce a distinct shifted dynamical wavevector $k_{+}$ for ultra-slow-roll.
Solving Eq. \eqref{eqn:USRMSequation}, then gives, for ultra-slow-roll
\begin{equation}
\label{eqn:USRtensormode}
    \begin{split}
        u_{k} & = \sqrt{\frac{\pi}{4}}\sqrt{3\eta_{t}-\eta}\Big[ C_{k}H_{\frac{3}{2}}^{(1)}(k_{+}(3\eta_{t}-\eta))\\
        &\hspace{2.2cm} + D_{k}H_{\frac{3}{2}}^{(2)}(k_{+}(3\eta_{t}-\eta))\Big] \, .
    \end{split}
\end{equation}
Furthermore, we invoke continuity of $u_{k}$ and $u_{k}'$ at $\eta = \eta_{t}$ and match Eqs. \eqref{eqn:KDtensormode} , \eqref{eqn:KDinitialconditions} and \eqref{eqn:USRtensormode}, we obtain the ultra-slow-roll mode coefficients of $u_{k}$ as
\begin{align}
    C_{k} &= \frac{i\pi\eta_{t}}{2\sqrt{2}}
    \Big[ k_{+}H^{(2)}_{0}(k_{-}\eta_{t})H^{(2)}_{1/2}(2k_{+}\eta_{t}) \nonumber\\
    &\quad - k_{-}H^{(2)}_{1}(k_{-}\eta_{t})H^{(2)}_{3/2}(2k_{+}\eta_{t})\Big] \label{eqn:USRCoefficientC} \, ,\\
    D_{k} &= \frac{i\pi\eta_{t}}{2\sqrt{2}}
    \Big[k_{-}H^{(2)}_{1}(k_{-}\eta_{t})H^{(1)}_{3/2}(2k_{+}\eta_{t}) \nonumber\\
    &\quad - k_{+}H^{(2)}_{0}(k_{-}\eta_{t})H^{(1)}_{1/2}(2k_{+}\eta_{t})\Big] \, , \label{eqn:USRCoefficientD}
\end{align}
where in the limit of zero curvature ($K =0$), we have $k_{-}^{2} \to \mathcal{K}^{2} \to k^{2}$ and $k_{+}^{2} \to \mathcal{K}^{2} \to k^{2}$. This picture is expectedly analogous to the case of curvature perturbations studied in~\cite{Contaldi:2003zv,Thavanesan:2020lov,Gessey-Jones:2021yky,Shumaylov:2021qje,Dineen:2023nbt}.

\subsection{Primordial tensor power spectrum}\label{sec:powerspectrumderivation}
With these solutions, we extend the analysis of \citet{Thavanesan:2020lov} to the case for tensor perturbations. The primordial tensor power spectrum of the metric perturbation $h$ then given by 
\begin{align}
    \mathcal{P}_{T}(k) &= 2 \left( \frac{k^{3}}{2\pi^{2}} \right) |h_{k}|^{2}\nonumber \\
    & \to \lim_{\eta \to 3\eta_{t}} 2 \left( \frac{k^{3}}{2\pi^{2}} \right) \Bigg| \frac{u_{k}}{a(\eta)}\Bigg| \\ 
    &=\frac{1}{2\pi^{2}a^{2}(3\eta_{t}-\eta)^{2}}\frac{k^{3}}{k_{+}^{3}}|C_{k} - D_{k}|^{2} \\
    & = A_{T}\frac{k^{3}}{k_{+}^{3}}|C_{k} - D_{k}|^{2} \, , \label{eqn:Powerspectrumlim}
\end{align}
where we have used the fact that $h_{k} = u_{k}/a(\eta)$, and absorbed the transition parameter $\eta_{t}$ (along with any formal divergences) into the tensor power spectrum amplitude $A_{T}$. On sub-transition scales, $k_{-} \to k_{+} \to k \gg 1/\eta_{t}$, we recover the nearly scale-invariant tensor spectrum --- the matching coefficients approach their Bunch–Davies values and the curvature shift is negligible
\begin{equation}
 \label{eqn:BunchDavies}
    |C_{k}| \simeq 1 \, , \quad |D_{k}| \ll  |C_{k}| \, , \quad \mathcal{P}_{T} \simeq A_{T} \, .
\end{equation}
Since we are working in the ultra-slow-roll regime, the leading order spectrum has no spectral tilt $n_{t}$, consistent with~\cite{Thavanesan:2020lov}. Although higher order terms can recover the tilt, we instead incorporate it phenomenologically by promoting the amplitude to a tilted form, $A_{T} \to A_{T}(k/k_{*})^{n_{t}} $.

\begin{figure*}[h!]
    \centering
    {\includegraphics [width=1\textwidth] {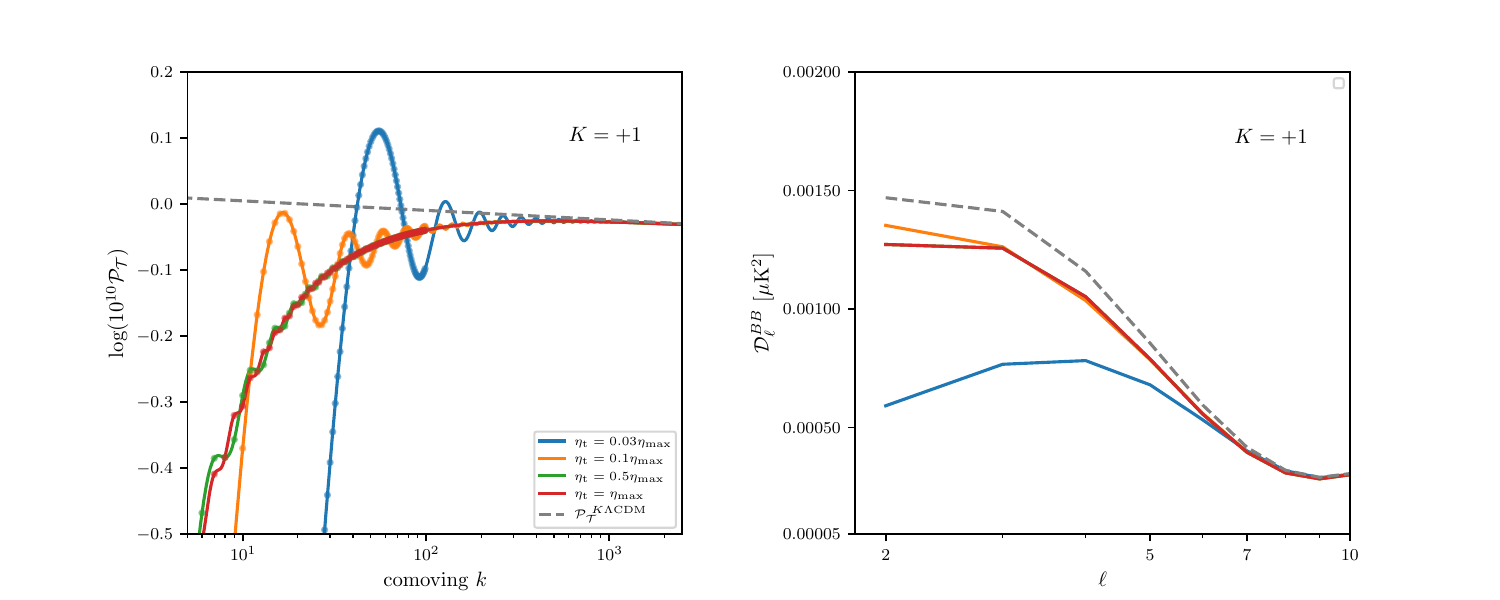}}
    \caption{Left: Primordial tensor power spectrum $\mathcal{P}_{\mathcal{T}}$ corresponding to the range of allowed values of the transition time $\eta_{t} $ for closed universes $K = +1$. Oscillations and a generic suppression of power are visible at low-$k$. Only integer values of comoving $k$ with $k \geq 3$ are allowed hence the jagged spectra, here the dots indicate the first $100$ comoving $k$ and for clarity we include the continuous spectrum. Right: the corresponding low-$\ell$ effects on the CMB unlensed $B$-mode power spectrum. The power law $K\Lambda$CDM  spectrum is highlighted in dashed grey. There is no appreciable deviation from the traditional power spectrum at higher $k$ and $\ell$ values.}
    \label{fig:ClosedTensorSpectra}
\end{figure*}

\begin{figure*}[h!]
    \centering
    {\includegraphics [width=1\textwidth] {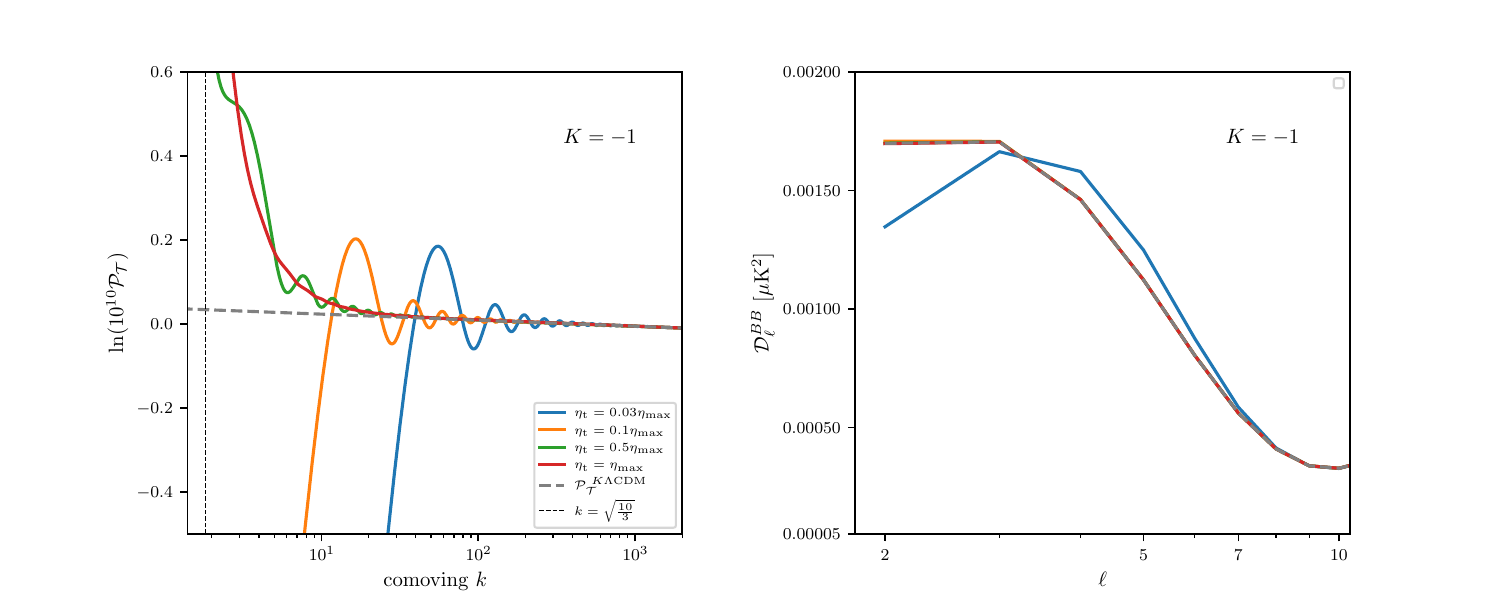}}
    \caption{Same as Figure \ref{fig:ClosedTensorSpectra}, but for open universes, $K = -1$. The primordial tensor power spectra exhibit a mild enhancement of power at low $k$, in contrast to the suppression observed in the closed case, notably dependent on the choice of transition time $\eta_{t}$. As $\eta_t \to \eta_{\text{max}}$, the leading order term in $\mathcal{P}_{\mathcal{T}}$ is an exponential, hence the exponential growth observed. At $k=\sqrt{10/3}$, $k_{-}$ becomes imaginary and introduces a cutoff to the spectrum. Right: The corresponding CMB $B$-mode spectra display a small excess of power at the low $\ell$, while converging to the standard $K\Lambda$CDM prediction at higher $\ell$. }
    \label{fig:OpenTensorSpectra}
\end{figure*}

Our analytical primordial tensor power spectrum, valid for each curvature $K \in \{+1,0,-1\}$ is thus parameterised by an amplitude $A_{T}$, spectral index $n_{t}$ and transition $\eta_{t}$
\begin{equation}
 \label{eqn:AnalyticalPT}
    \mathcal{P}_{T}(k) =  A_{T}\left(\frac{k}{k_{*}}\right)^{n_{t}}\frac{k^{3}}{k_{+}^{3}}|C_{k}(\eta_{t}) - D_{k}(\eta_{t})|^{2} \, ,
\end{equation}
where $C_{k}$ and $D_{k}$ are defined by Eqs.  \eqref{eqn:USRCoefficientC} and \eqref{eqn:USRCoefficientD}, and involve Hankel functions with curvature–shifted wavevectors $k_{\pm}$ defined in Eqs.  \eqref{eqn:KDMSequation} and \eqref{eqn:USRMSequation}. The resulting power spectra of $\mathcal{P}_\mathcal{T}$ are shown in Figures \ref{fig:ClosedTensorSpectra} and \ref{fig:OpenTensorSpectra}.\par
Combining our analytical tensor spectrum \eqref{eqn:AnalyticalPT} with the scalar template of Thavanesan et al. \cite{Thavanesan:2020lov}, we obtain an explicit expression for the tensor-to-scalar ratio,
\begin{equation}
    \label{eqn:tenstoscal}
    r(k) = \frac{A_{T}}{A_{S}}\left(\frac{k}{k_{*}}\right)^{n_t - (n_s - 1)}\left(\frac{q_{+}}{k_{+}}\right)^{3}\frac{|C_{k}(\eta_{t}) - D_{k}(\eta_{t})|^{2}}{|\alpha_{k}(\eta_{t}) - \beta_{k}(\eta_{t})|^{2}},
\end{equation}
where $q_{+}$ and $\alpha_k$, $\beta_k$ are the scalar-sector shifted wavevector and mode coefficients defined in \cite{Thavanesan:2020lov}.\par


We highlight that through the analytic framework used to derive the curvature-induced modifications for scalar perturbations, we are able to extend the formalism to the tensor sector and identify the corresponding dynamical effects of spatial curvature. \par 
In particular, Eqs. \eqref{eqn:KDMSequation} and \eqref{eqn:USRMSequation} constitute the central results of this work, demonstrating that spatial curvature manifests dynamically as a shift in the effective wavevector in both the kinetically dominated and ultra-slow roll regimes.

\section{Discussion}\label{sec:discussion}
The analytic construction of curved inflationary dynamics presented in Section \ref{sec:computingCurvedSpectra} isolates the effects of curvature on the primordial tensor power spectrum. Mathematically, these effects are attributed to shifts in the wavevector participating dynamically in Eqs. \eqref{eqn:KDMSequation} and \eqref{eqn:USRMSequation}. Just as with the scalar case \cite{Thavanesan:2020lov}, the Laplacian operator in \eqref{eqn:Tensormode}, in Fourier space, is replaced by a scalar wavevector shifted by a curvature term. The shifted wavevector introduces phase-driven oscillations in the primordial tensor power spectrum $\mathcal{P}_{T}(k)$ dependent on the transition time $\eta_{t}$. Increasing $\eta_{t}$ shifts oscillations to smaller $k$ and enhances their visibility, the oscillations seen in the numerically generated curved primordial power spectra for closed inflating universes \cite{Handley:2019} are here reproduced for the tensor power spectra.

Furthermore, following~\cite{Thavanesan:2020lov,Handley:2019}, our analytic construction reveals that in open universes, curvature once again manifests as a shift of the dynamical wavevector, producing the same phase-driven oscillatory features identified in the scalar sector. In closed universes, the constraint $k \in \mathbb{Z}{>2}$, below which the frequency of the oscillatory solutions becomes imaginary, signals a breakdown of the approximation at large $\eta{t}$, consistent with the behaviour observed for scalar perturbations. By contrast, in open universes the condition $k = \sqrt{10/3}$ causes $k_{-}$ to become imaginary, introducing a natural large-scale cutoff in the tensor spectrum — a distinctive feature absent in the scalar case.

We treat $K\Lambda$CDM as a one-parameter extension of $\Lambda$CDM in which spatial curvature, quantified by $\Omega_{K}$, is allowed to deviate from zero. Within $K\Lambda$CDM the primordial tensor power spectrum is taken to be nearly scale-invariant
\begin{equation}
    \label{eqn:KLCDMspectrum}
    \mathcal{P}_{T}(k) = A_{T}\left(\frac{k}{k_{*}}\right)^{n_{t}} \, ,
\end{equation}
where $k_{*}$ is the pivot scale at which the amplitude $A_{T}$ and the tilt $n_{t}$ of the primordial spectrum are evaluated. Using the \textit{Planck} 2018 results including CMB lensing (TTTEEE$+$lowl$+$lowE$+$lensing) data, the best fit parameter values for closed universes are $A_{s} = 2.0771 \pm 0.1017\times10^{-9}$ and $r_{k_*=0.01}< 0.066$, where we have chosen $r_{k_*=0.01} =0.05$ for this investigation.

Previous work by Handley \cite{Handley:2019} showed that including fully numerical treatments of curved universes, produces low-$k$ oscillations and suppression of power --- with no dependence on initial conditions, showing deviation from the supplemental scalar form of Eq. \eqref{eqn:AnalyticalPT}. Furthermore, that the $(k/k_{*})^{n_{t}}$ tilt is a higher-order effect tied to the specific slow-roll potential. Such corrections can be obtained by computing the higher order terms of the \textit{logo linear expansion} in Appendix \ref{sec:Appendix}.

As a pragmatic phenomenological template, we follow the procedure in \cite{Thavanesan:2020lov}. We rescale the normalised curvature spectrum $\mathcal{P}_{\mathcal{R}}$ with the best-fit scalar power spectrum amplitude $A_{s}$ and manually insert the tilt, then convert these results to the tensor $\mathcal{P}_{\mathcal{T}}$ with the standard relation \cite{Starobinsky:1979ty,Guth:1980zm,Linde:1981mu,Starobinsky:1980te}.

We present the analytical
primordial power spectrum $\mathcal{P}_{\mathcal{T}}$ for a range of transition times $\eta_{t}$ in closed universes ($K =+1$) in Figure \ref{fig:ClosedTensorSpectra} and propagate these templates through to the unlensed $B$-mode CMB anisotropy power spectrum \cite{CLASS}. The same is done for open universes ($K=-1$) in Figure \ref{fig:OpenTensorSpectra}. The CMB spectra are computed with parameter values consistent with the best-fit data for each curvature scenario. We use the \textit{Planck} 2018 TTTEEE+lowl+lowE+lensing best-fit parameters for the closed case. Whereas for the open case we calculate the mean posterior derived from lensing data under the constraint $\Omega_{K} >0$ (i.e. $K = -1$) using the \textbf{anesthetic} package \cite{Handley:2019mfs}.

Solving the horizon problem requires that the amount of conformal time during inflation $\eta_{i}$ is separately greater that the amount of conformal time before and after $\eta_{t}$. In our setup, the ultra-slow-roll branch of Eq.~\eqref{eqn:SFcases} yields $\eta_{i} = 2\eta_{t}$. automatically ensuring that $\eta_{i}> \eta_{t}$  for the pre-inflationary kinetic dominance regime. Consistently also imposing a post transition bound $\eta_{+} < 2\eta_{t}$. The implications of these constraints for exact numerical solutions are discussed in detail by~\cite{Handley:2019}.

Figures \ref{fig:ClosedTensorSpectra} and \ref{fig:OpenTensorSpectra} show the behavior of the computed $\mathcal{P}_{\mathcal{T}}$ spectra for different values of the transition time $\eta_{t}$. Adjusting $\eta_{t}$ shifts the cutoff scale and changes the suppression of power as well as the oscillatory features, as with the scalar case \cite{Lasenby:2003ur}. In closed universes ($K = +1$), the depth of the low-$k$ suppression increases with $\eta_{t}$. By contrast for sufficiently large $\eta_{t}$ open universes ($K=-1$) exhibit a runoff enhancement of power.

Figures \ref{fig:ClosedPolarisation} and \ref{fig:OpenPolarisation} illustrate the corresponding polarisation pattern in the Cosmic Microwave Background (CMB) anisotropies sourced by tensor perturbations in curved inflating universes. The figures show the predicted large-scale $B$-mode polarisation pattern arising from the primordial tensor modes derived in this work, superimposed on the temperature anisotropy field \cite{Preece:2014qaa}. Additionally Figures \ref{fig:ClosedPolarisationDifference} and \ref{fig:OpenPolarisationDifference} highlight the residual differences between the K$\Lambda$CDM case and those parametrised by a given transition time value $\eta_{t}$. Making explicit the curvature-induced features in the $B$-mode signal.

We here show that the analytic approximations of \cite{Handley:2019,Thavanesan:2020lov} extend naturally to primordial tensor spectra in curved universes. In doing so we demonstrate the framework in \cite{Thavanesan:2020lov} can isolate curvature effects on tensor perturbations without introducing additional model dependence, providing a clean, potential-agnostic template for curved inflationary dynamics.

\section{Conclusion}\label{sec:conclusion}

The inflationary scenario provides a compelling mechanism for explaining the observed flatness, homogeneity, and isotropy of the universe, but offers limited insight into the pre-inflationary state that set the initial conditions for the inflationary phase. A complete understanding of inflation therefore requires treating the initial spatial curvature as an independent degree of freedom rather than assuming an exactly flat initial hypersurface. Even small deviations from flatness can leave distinct imprints on the primordial perturbation spectra, motivating the analytic treatment of curved inflationary universes.

In this work, we have extended the analytic framework developed in~\citet{Thavanesan:2020lov} to the tensor sector, deriving compact analytical approximations for the primordial tensor power spectrum in generally curved inflationary universes. We reformulate the tensor-mode equation of motion in conformal time, explicitly incorporating the effects of curvature through shifts in the dynamically relevant wavevector. This reformulation reveals that curvature acts as a systematic modification of the tensor mode dynamics, introducing characteristic oscillations and power suppression at large scales in the primordial tensor spectrum $\mathcal{P}_{\mathcal{T}}(k)$. 

Our analytic templates capture the essential features observed in full numerical solutions, providing physical insight into how curvature influences the propagation and freezing of tensor modes. In particular, the oscillatory features and low-$\ell$ cut-offs that appear in the scalar spectrum have direct tensor analogues, controlled by the transition time $\eta_t$ between the kinetically dominated and inflationary phases. Varying $\eta_t$ shifts the position and amplitude of these features, offering a physically transparent parameterisation of curvature effects. For closed universes ($K=+1$), we find a suppression of power and oscillatory behaviour at low $k$, while open universes ($K=-1$) display an enhancement of large-scale power for sufficiently large transition times. These effects persist when the spectra are propagated to the BB power spectrum of the CMB, suggesting that curvature-induced modifications could be detectable in forthcoming polarisation measurements.

Our analysis also confirms that curvature manifests dynamically as a shifted effective wavevector in both open and closed cases, consistent with the behaviour of scalar modes identified in~\citet{Thavanesan:2020lov}. The analytic form of our spectra provides a clear interpretation of this behaviour, showing that curvature modifies the oscillatory phase structure of the solutions rather than introducing fundamentally new dynamical degrees of freedom. This phase-driven interpretation unifies the scalar and tensor analyses and clarifies how curvature systematically influences the inflationary initial conditions.

Phenomenologically, our results indicate that even small departures from flatness can leave observable imprints on the tensor power spectrum, with corresponding effects on the large-angle $B$-mode polarisation of the CMB. Upcoming experiments such as the \emph{Simons Observatory}~\cite{SimonsObservatory:2018koc} and \emph{CMB-S4}~\cite{CMB-S4:2020lpa} will probe these scales with unprecedented precision, potentially allowing the first direct constraints on primordial curvature through tensor modes.

More broadly, the potential-independent analytic framework developed here provides a versatile tool for exploring curvature effects across a range of inflationary models. By isolating universal geometric contributions, it offers a bridge between numerical analyses and physical interpretation, complementing full simulations and enabling model-independent forecasts. Future work could extend this framework to include higher-order corrections and mixed scalar–tensor correlations, providing a unified analytic description of curvature signatures in the early universe.

In summary, the analytic approach presented here demonstrates that curvature leaves a distinctive and physically interpretable imprint on the primordial tensor spectrum. These results extend and unify previous scalar analyses, offering a coherent framework through which to connect inflationary curvature with forthcoming observational probes of primordial gravitational waves.

\section*{Acknowledgements}
The authors thank Aron Clark Wall for providing helpful comments on earlier drafts of the paper. AT was supported in part by the Heising-Simons Foundation, the Simons Foundation, the Bell Burnell Graduate Scholarship Fund, the Cavendish (University of Cambridge) and a KITP Graduate fellowship. AT was also supported by grant no. NSF PHY-2309135 to the Kavli Institute for Theoretical Physics (KITP). AT also acknowledges support from the SNF starting grant “The Fundamental Description of the Expanding Universe". AT also gratefully acknowledges hospitality from the Flatiron Institute, which enabled the start of this collaboration, as well as the Perimeter Institute, while working on this article. For the purpose of open access, we have applied a CC BY public copyright licence to any Author Accepted Manuscript version arising.

\appendix
\section{Logolinear expansions in conformal time}\label{sec:Appendix}
Logolinear series expansions~\cite{Handley:2019bzs} for a general function $x(\eta)$ are given as
\begin{equation}
    x(\eta) = \sum_{j,k} [x^k_j] \: \eta^j {\left( \log \eta \right)}^k \, ,
    \label{eqn:logolineardefinition}
\end{equation}
where $[x^k_j]$ are twice-indexed real constants which define the series. The square brackets here are used to separate powers from superscript. To derive the logolinear expansions for the kinetic dominance regime, we begin by restating Eqs. \eqref{eqn:KleinGordon2} and \eqref{eqn:HubbleDerivative} as
\begin{align}
    \dderiv{N} +\frac{1}{3}\left( {\deriv{\phi}}^2 - a^2V(\phi) \right) &=0 \, ,
    \label{eqn:raychaudhuri_N}\\
    \dderiv{\phi} + 2 \deriv{N}\deriv{\phi} + a^2\frac{\d{}}{\d{\phi}}V(\phi) &=0 \, .
    \label{eqn:klein_gordon_N}
\end{align}
Here $N = \log a$ has been used rather than $\conformalH$ such that the differential equations now appear to second order, which we can then in turn convert to a first order system of equations
\begin{align}
    \dot{N} &= h \, ,
    \qquad 
    \dot{\phi} = v \, ,
    \nonumber\\
    \dot{h} &= h -\frac{1}{3} v^2 + a^2\frac{1}{3} \eta^2V(\phi) \, ,
    \nonumber\\
    \dot{v} &= v - 2 v h - a^2\eta^2\frac{\d{}}{\d{\phi}}V(\phi) \, ,
    \label{eqn:dsys}
\end{align}
where dots indicate derivatives with respect to logarithmic conformal time $\log\eta$, such that $\dot{x}=\frac{\d{}}{\d{\log \eta}} x$.

To analytically determine approximate solutions for curved cosmologies we will consider series expansions for a general function $x(\eta)$ of the form 
\begin{equation}
    x(\eta) = \sum_j x_j(\eta)\: \eta^j \quad\Rightarrow\quad \dot{x}(\eta) = \sum_j (\dot{x}_j + j x_j)\: \eta^j \, .
    \label{eqn:logolinear}
\end{equation}

Substituting in our series definition from \cref{eqn:logolinear} and equating coefficients of $\eta^j$, we find that \cref{eqn:dsys} becomes
\begin{align}
    \dot{N}_j + j N_j &= h_j \, ,
    \qquad
    \dot{\phi}_j + j \phi_j = v_j \, ,
    \nonumber\\
    \dot{h}_j + j h_j &= h_j + \frac{1}{3} V(\phi)e^{2N_\p}e^{\sum_{q>0}N_q(\eta)\eta^q}|_{j-3} -\smashoperator{\sum_{p+q=j}}\frac{v_p v_q}{3} \, ,
    \nonumber\\
    \dot{v}_j + j v_j &= v_j - \frac{\d{V(\phi)}}{\d{\phi}}e^{2N_\p}e^{\sum_{q>0}N_q(\eta)\eta^q}|_{j-3} - 2 \smashoperator{\sum_{p+q=j}} v_p h_q \, ,
    \label{eqn:dsysj}
\end{align}
where exponentiation of logolinear series was discussed in~\cite{Handley:2019bzs}, to better interpret the behavior of solutions to the differential equations.
One can also consider the equivalent of \eqref{eqn:Friedmann1}
\begin{equation}
    \frac{1}{3}V(\phi)e^{2N_\p}e^{\sum_{q>0}N_q(\eta)\eta^q}|_{j-3} +\smashoperator{\sum_{p+q=j}}{\frac{1}{6}}v_p v_q - h_p h_q =K|_{j-2} \, .
    \label{eqn:friedmann_logt}
\end{equation}

We first solve for the $j=0$ case of \cref{eqn:dsysj}, substituting that value for $j$ shows that it is equivalent to ~\cref{eqn:dsys} with $V=0$. Thus one can solve \cref{eqn:dsysj} using kinetically dominated solutions
\begin{align}
    N_0 &= N_\p + \frac{1}{2}\log \eta \, , &h_0 &= \frac{1}{2} \, , \nonumber\\
    \phi_0 &= \phi_\p \pm \sqrt{\frac{3}{2}}\log \eta \, , &v_0 &=\pm\sqrt{\frac{3}{2}} \, ,
    \label{eqn:0_sol}
\end{align}
where $N_p$ and $\phi_p$ are constants of integration. We expect there to be four constants of integration {\em a priori}. One such constant is fixed by placing the singularity at $\eta=0$, while the other is implicitly set by the curvature parameter. Hence \cref{eqn:0_sol} provides a complete solution to $j=0$ only in the flat case ($K=0$). Nevertheless, we may still adopt \cref{eqn:0_sol} to be the base term of the logolinear expansion. The final constant of integration thus naturally emerges through the inclusion of higher order corrections.

We now consider $j\ne 0$, where \cref{eqn:dsysj} can be written as a first order linear inhomogeneous vector differential equation
\begin{equation}
    \dot{x}_j + A_j x_j = F_j \, ,
    \label{eqn:linear_master}
\end{equation}
where $x=(N,\phi,h,v)$, and $A_j$ is a (constant) matrix given as
\begin{align}
    A_j &= \left(
    \begin{array}{cccc}
        j & 0 & -1 & 0 \\
        0 & j & 0 & -1 \\
        0 & 0 & j-1 & \frac{2}{3}v_0 \\
        0 & 0 & 2v_0 & j-1+2h_0 \\
    \end{array}
    \right) \, ,\nonumber\\
    &= \left(%
    \begin{array}{cccc}
        j & 0 & -1 & 0 \\
        0 & j & 0 & -1 \\
        0 & 0 & j-1 & \pm\sqrt{\frac{2}{3}} \\
        0 & 0 & \pm\sqrt{6} & j \\
    \end{array}
    \right) \, ,\label{eqn:A}
\end{align}
and $F_j$ represents a vector polynomial in $\log \eta$ depending only on the preceding series terms $x_{p<j}$
\begin{align}
    F_j &=
    \left(
    \begin{array}{c}
        0\\
        0\\
        \frac{1}{3} V(\phi)e^{2N_\p}e^{\sum_{q>0}N_q(\eta)\eta^q}|_{j-3}-\smashoperator{\sum_{\substack{p+q=j\\p\ne j,q\ne j}}} \frac{1}{3}v_p v_q \\
        - \frac{\d{V(\phi)}}{\d{\phi}}e^{2N_\p}e^{\sum_{q>0}N_q(\eta)\eta^q}|_{j-3} - 2 \smashoperator{\sum_{\substack{p+q=j\\p\ne j,q\ne j}}} v_p h_q 
    \end{array}
    \right) \, .\label{eqn:Fj}
\end{align}

As with \cite{Thavanesan:2020lov}, the linear differential \cref{eqn:linear_master} can be expressed as the sum of a complementary function ($x_j^\mathrm{cf}$) containing four free parameters and a particular integral ($x_j^\mathrm{pi}$), at each $j$, such that $x_j = x_j^\mathrm{cf} + x_j^\mathrm{pi}$. The free parameters correspond to the degrees of gauge freedom previously discussed in~\citep{Handley:2019bzs}.

The homogeneous version of \cref{eqn:linear_master} can be solved exactly, as $A_j$ is a constant matrix
\begin{equation}
 \label{eqn:homogeneous_sol}
    \frac{\d{x_j^\mathrm{cf}}}{\d{\log \eta}} + A_j x_j^\mathrm{cf} = 0  \quad\Rightarrow\quad x_j^\mathrm{cf} = e^{-A_j\log \eta}[x_j^0] \, ,
\end{equation}
where $[x_j^0]$ is a constant vector parameterizing initial conditions.
The matrix exponential is derived by firstly computing the eigenvectors and eigenvalues of $A_j$
\begin{align}
    e_\sft &= \left(%
    \begin{array}{cccc}
        1& \pm\sqrt{6} & \frac{(\sqrt{6}-18)}{12}& \mp\sqrt{6} \\
    \end{array}
    \right) \, ,& 
    A_j e_\sft{} &= (j+1)\cdot e_\sft \, ,
    \nonumber\\
    e_b &= \left(%
    \begin{array}{cccc}
        1& \mp\frac{\sqrt{6}}{2} & \frac{(\sqrt{6}+18)}{12}& \mp\sqrt{6} \\
    \end{array}
    \right) \, ,&
    A_j e_b &= (j-2)\cdot e_b,
    \nonumber\\
    e_n &= \left(%
    \begin{array}{cccc}
        1& 0& 0& 0\\
    \end{array}
    \right) \, ,&
    A_j e_n &= j\cdot e_n \, ,
    \nonumber\\
    e_\phi{} &= \left(%
    \begin{array}{cccc}
        0& 1& 0& 0\\
    \end{array}
    \right) \, ,&
    A_j e_\phi{} &= j\cdot e_\phi \, .\label{eqn:eigenvalues}
\end{align}

By expressing the initial conditions $[x_j^0]$ in terms of the eigenbasis given in \cref{eqn:eigenvalues} with parameters $\tilde{N},\tilde{\phi},\tilde{b}, \tilde{\sft}$, we obtain
\begin{align}
    x_j^\mathrm{cf}
    &= e^{-A_j\log \eta}(\tilde{N}e_n + \tilde{\phi}e_\phi + \tilde{b} e_b + \tilde{\sft} e_\sft)\nonumber\\
    &= \left(\tilde{N}e_n + \tilde{\phi}e_\phi + \tilde{b}e_b \eta^{2} + \tilde{\sft} e_\sft \eta^{-1}\right)\eta^{-j} \, .
    \label{eqn:complementary_function}
\end{align}

The parameters $\tilde{N}$ and $\tilde{\phi}$ can be absorbed into the definitions of $N_\p$ and $\phi_\p$ respectively. Setting $\tilde{\sft}=0$ effectively places the singularity to be at $\eta=0$ as an initial condition without loss of generality, since this terms diverges more rapidly than leading terms as $\eta \to 0$. The remaining undetermined integration constant $\tilde{b}$, corresponds to the constant missing from \cref{eqn:0_sol}, this constant is determined by the spatial curvature of the universe via \cref{eqn:friedmann_logt}
\begin{equation}
    \tilde{b} = -\tfrac{1}{3}K \, .
    \label{eqn:curvature_relation}
\end{equation}

 We can now express the curvature parameter $K$ in terms of $\tilde{b}$ through this relation. In the analysis presented in the main body of the paper we omit $\tilde{b}$ and explicitly denote curvature terms with $K$ in the series solutions via \cref{eqn:curvature_relation}. From \eqref{eqn:complementary_function} it is evident that the first order curvature correction terms depend scale as $\eta^2$. Solving to first order in $K$ then yields the following parameter definitions.
 \begin{align}
    N_1 &= N_\p + \frac{1}{2}\log \eta  + \frac{K}{3}\eta^{2} \, ,\\
    \phi_1 &= \phi_\p \pm \sqrt{\frac{3}{2}}\log \eta \pm \frac{\sqrt{6}K}{3}\eta \, . 
    \label{eqn:1_sol}
\end{align}
We must now determine a particular integral of \cref{eqn:linear_master}, given the known form of $F_j$ at each recursive step.
Assuming a trial solution in the form of $x_j(\eta)=\sum_{k=0}^{N_j} [x^k_j] {(\log \eta)}^k$, and expanding ${F_j =\sum_{k=0}^{N_j} [F^k_j] {(\log t)}^k}$, we can equate coefficients of equal powers of $\log \eta$ to obtain
\begin{equation}
    (k+1)[x^{k+1}_j] + A_j [x^k_j] = [F^k_j] \, ,
    \label{eqn:linear_master_no_j}
\end{equation}
which leads to a descending recursion in $k$
\begin{equation}
    [x^{N_j+1}_k]=0,\quad [x^{k-1}_j] = A_j^{-1}( [F_j^{k-1}]  - k [x^{k}_j]) \, .
    \label{eqn:recursion_relation}
\end{equation}

Note that the recursion relation in \cref{eqn:recursion_relation} breaks down whenever $A_j$ becomes non-invertible--that is, when any of the eigenvalues in \cref{eqn:eigenvalues} vanish ($j=-1,0,2$). In these cases, the system is underdetermined and admits an infinite family of solutions parameterised along the corresponding eigenvector directions. These apparent degeneracies can be consistently absorbed into a corresponding constant of integration. In the main body of this paper we include curvature terms to order $K^{2}$, corresponding to obtaining solutions up to $j=4$. Found by introducing 
\begin{equation}
 \label{eqn:j=4_FandA}
    A_{4} = 
    \begin{pmatrix}
        4 & 0 & -1 & 0 \\
        0 & 4 & 0 & -1 \\
        0 & 0 & 3 & \frac{2}{3}v_{0} \\
        0 & 0 & 2v_{0} &  4 \\
    \end{pmatrix} \, ,  \;\;\; F_{4} = 
    \begin{pmatrix}
        0  \\
        0 \\
        -\frac{2}{9}K^{2} \\
        \pm \frac{4\sqrt{6}}{9}K^{2} \\
    \end{pmatrix} \, .
\end{equation}

Redefining the base of our recursion relation in \cref{eqn:recursion_relation}, would generate an infinite series. However, all but a finite number of terms would merely contribute to a redefinition of constants of integration, or introducing a non-zero $\tilde{\sft}$. The latter is excluded since it would shift the singularity to a non-zero conformal time $\eta \ne 0$, which would contradict our initial condition.

\section{Simulated $B$-mode Polarisation Plots}\label{sec:AppendixB}
We present the simulated $B$-mode polarisation taken from the closed ($K=+1$) and open ($K=-1$) spectra in Figures \ref{fig:ClosedTensorSpectra} and \ref{fig:OpenTensorSpectra}. The CMB anisotropy are seen at order $(10^{-5}\mu K^2)$ and thus we also plot the residual between each cosmological scenario parametrised by $\eta_t$ and the reference $K\Lambda$CDM spectrum. This comparison allows us to isolate the effects of varying the value of $\eta_t$ to the $B$-mode signals.
\begin{figure*}[t!]
\captionsetup[subfigure]{labelformat=empty}
\centering
\subfloat[$\eta_{t}= 0.03\eta_{\text{max}}$\label{fig:BmodeClosed03}]{
    \includegraphics[width=0.5\textwidth]{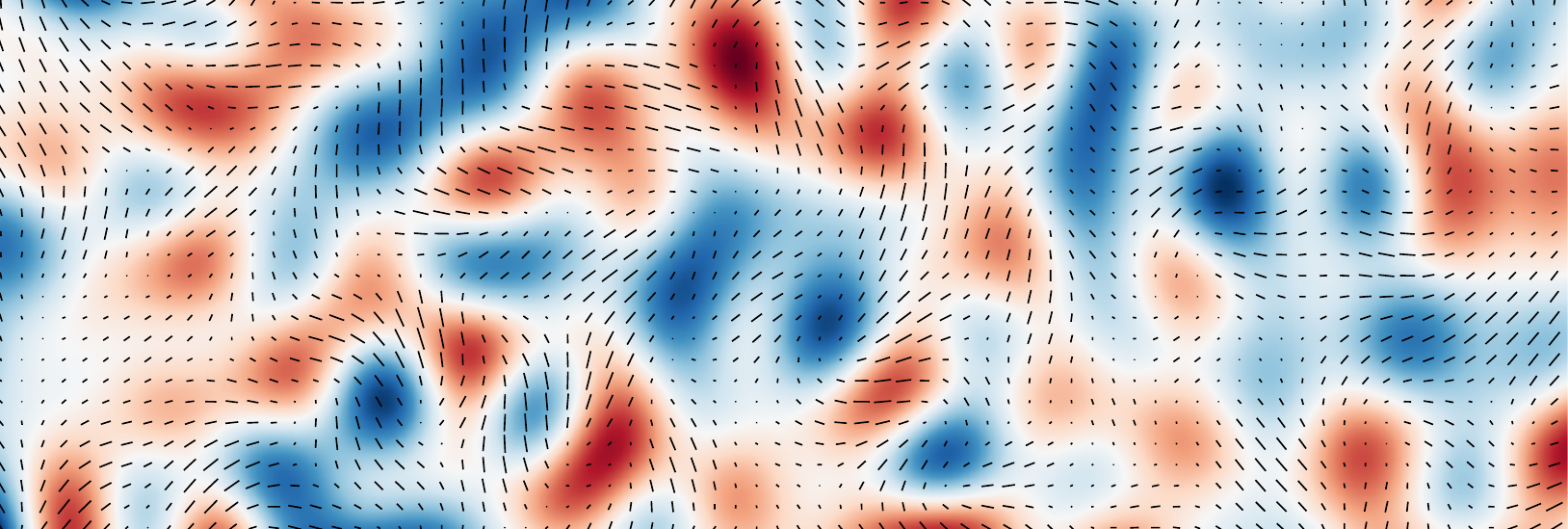}
}
\subfloat[$\eta_{t}= 0.1\eta_{\text{max}}$\label{fig:BmodeClosed1}]{
    \includegraphics[width=0.5\textwidth]{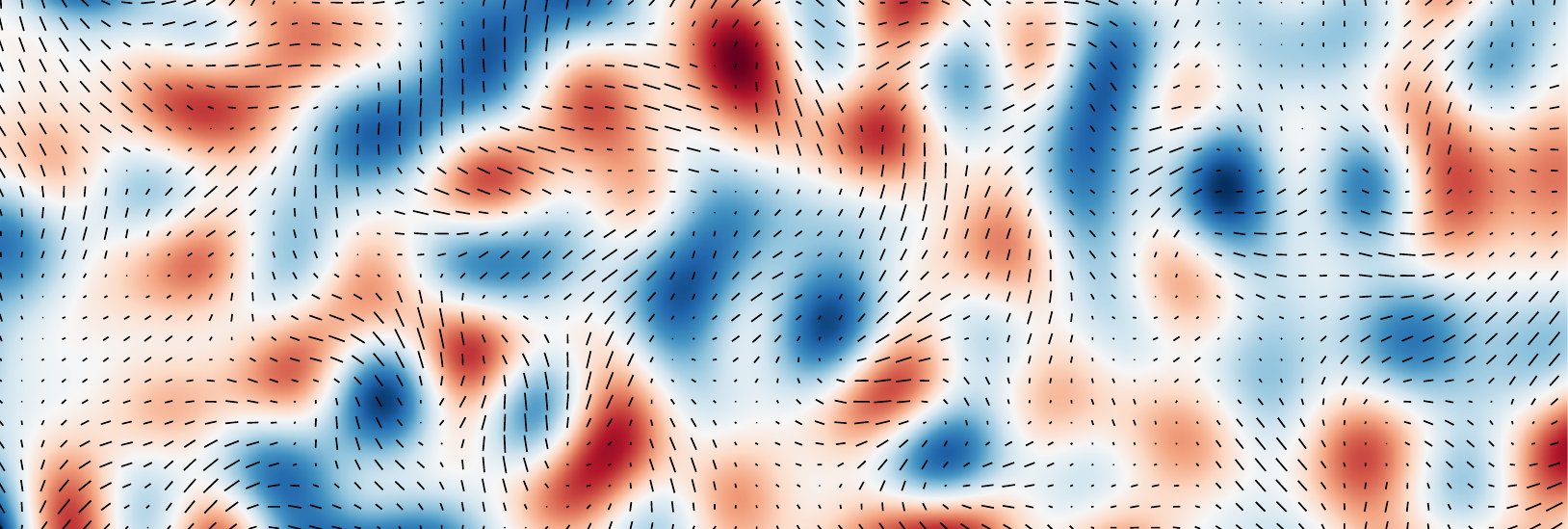}
} \\[1em] 
\subfloat[$\eta_{t}= 0.5\eta_{\text{max}}$\label{fig:BmodeClosed5}]{
    \includegraphics[width=0.5\textwidth]{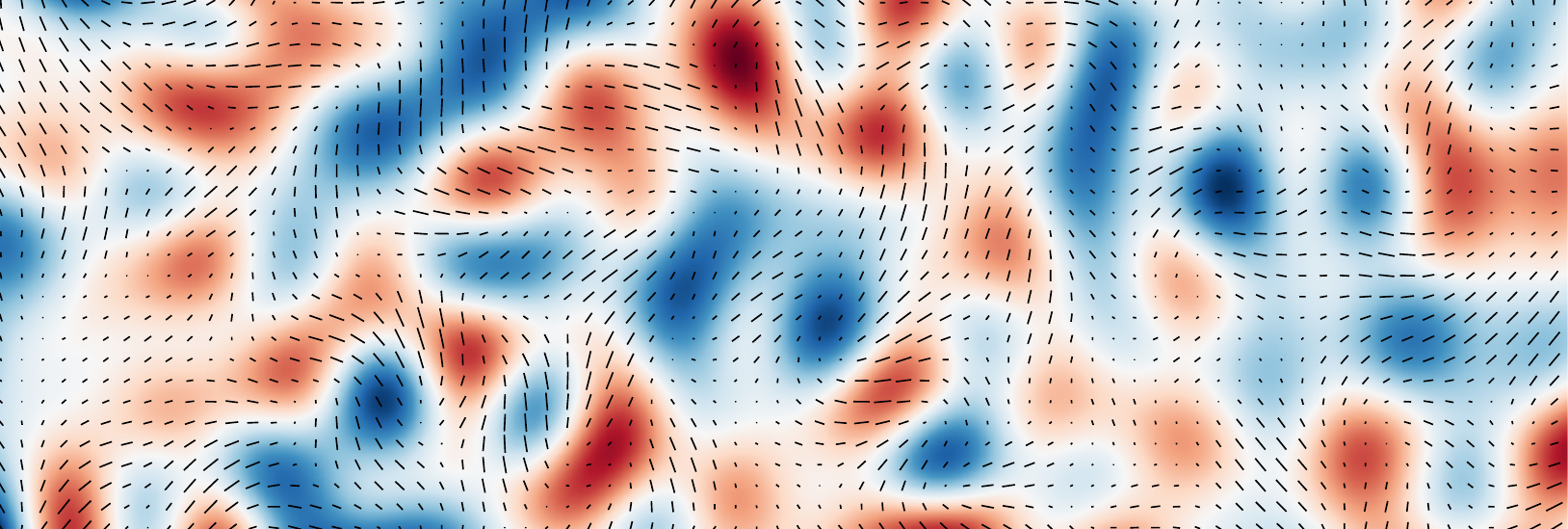}
}
\subfloat[$\eta_{t}=\eta_{\text{max}}$\label{fig:BmodeClosedMax}]{
    \includegraphics[width=0.5\textwidth]{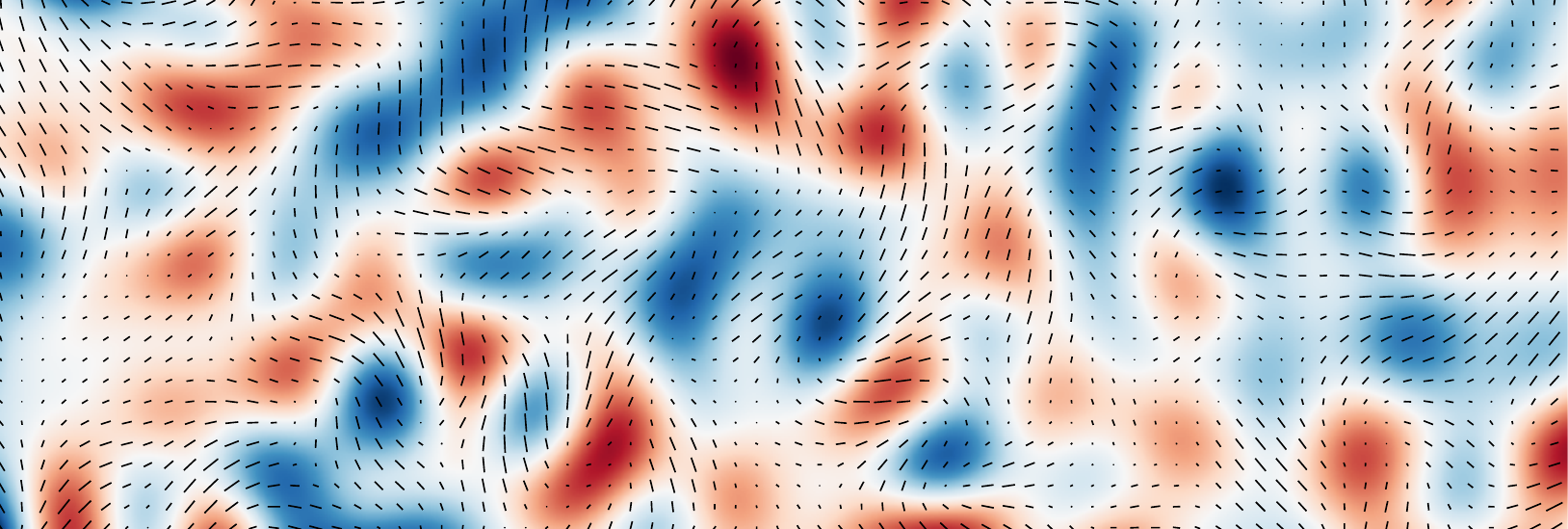}
} \\[1em] 
\subfloat[$K\Lambda$CDM\label{fig:KLCDM}]{
    \includegraphics[width=0.5\textwidth]{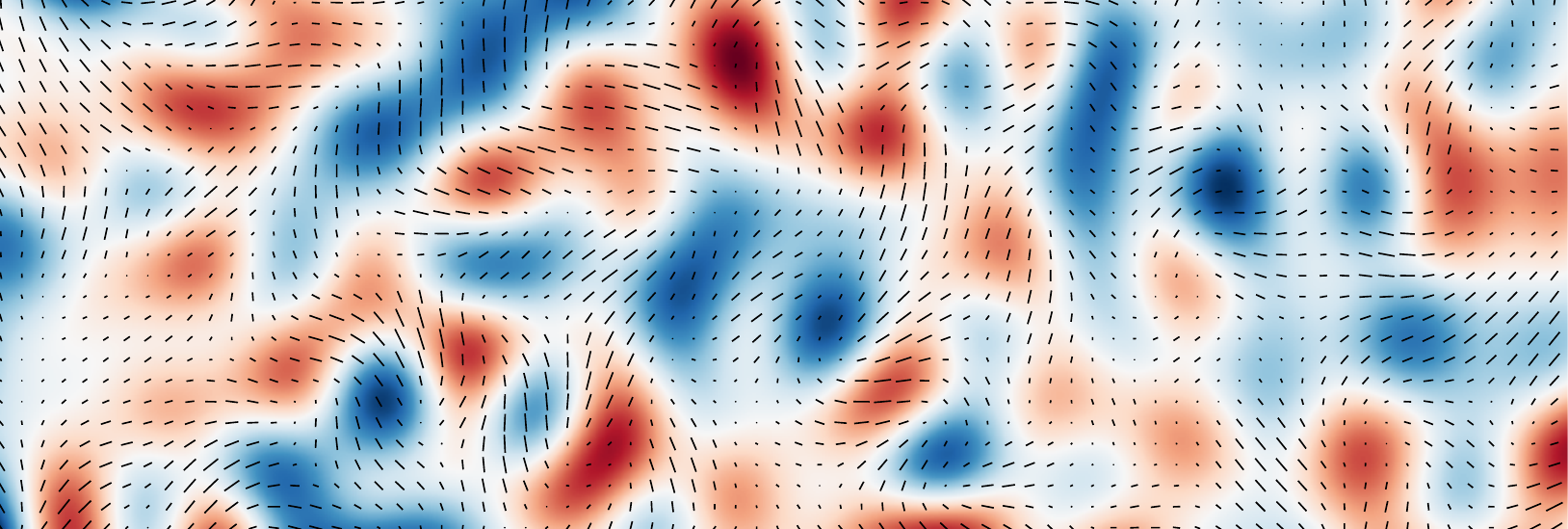}
}
\caption{\label{fig:ClosedPolarisation} Simulated closed universes $(K=+1)$ CMB $B$-mode polarisation maps of four chosen values of transition time $\eta_t \in \{(0.03,0.1,0.5,1)\times \eta_{\text{max}} \}$ where $\eta_{\text{max}} = \pi/4$, and one for fiducial closed $K\Lambda$CDM. Each image is a $15^{\circ}\times 5^{\circ}$ flat-sky projection capturing the low multipole ($< 500$) signatures of the $B$-mode polarisation. The polarisation maps have the colourbar truncated to $\pm7\times 10^{-5} \mu K^2$.}
\end{figure*}

\begin{figure*}[t!]
\captionsetup[subfigure]{labelformat=empty}
\centering
\subfloat[$\eta_{t}= 0.03\eta_{\text{max}}$\label{fig:BmodeOpen03}]{
    \includegraphics[width=0.5\textwidth]{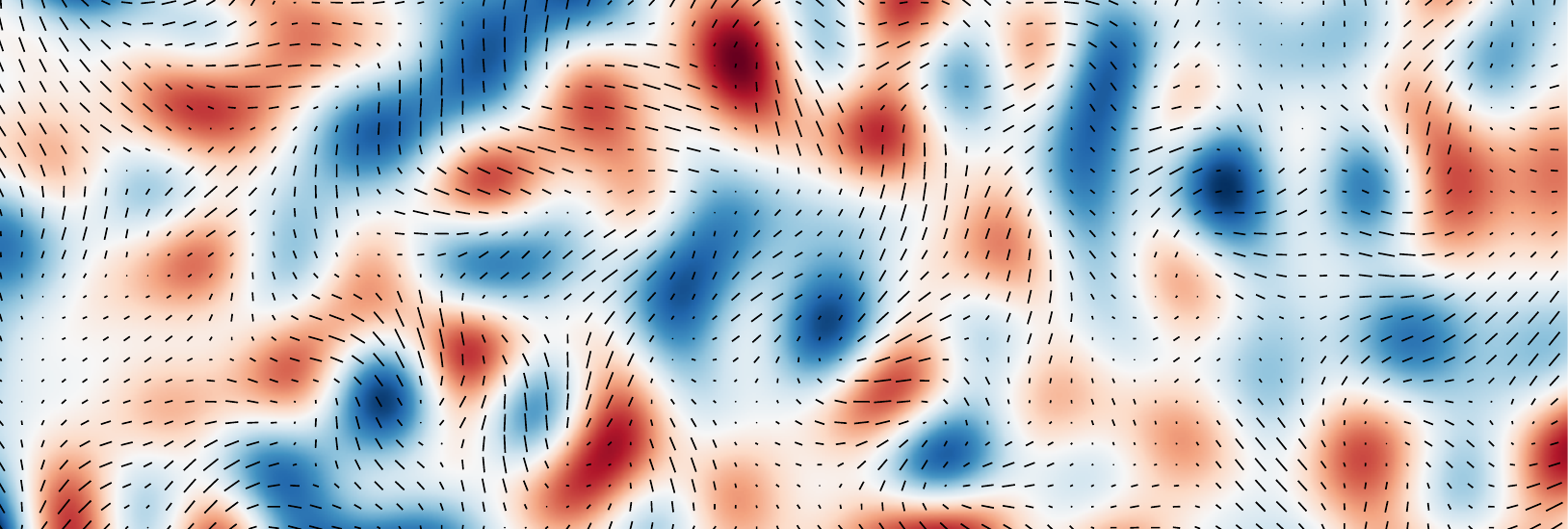}
}
\subfloat[$\eta_{t}= 0.1\eta_{\text{max}}$\label{fig:BmodeOpen1}]{
    \includegraphics[width=0.5\textwidth]{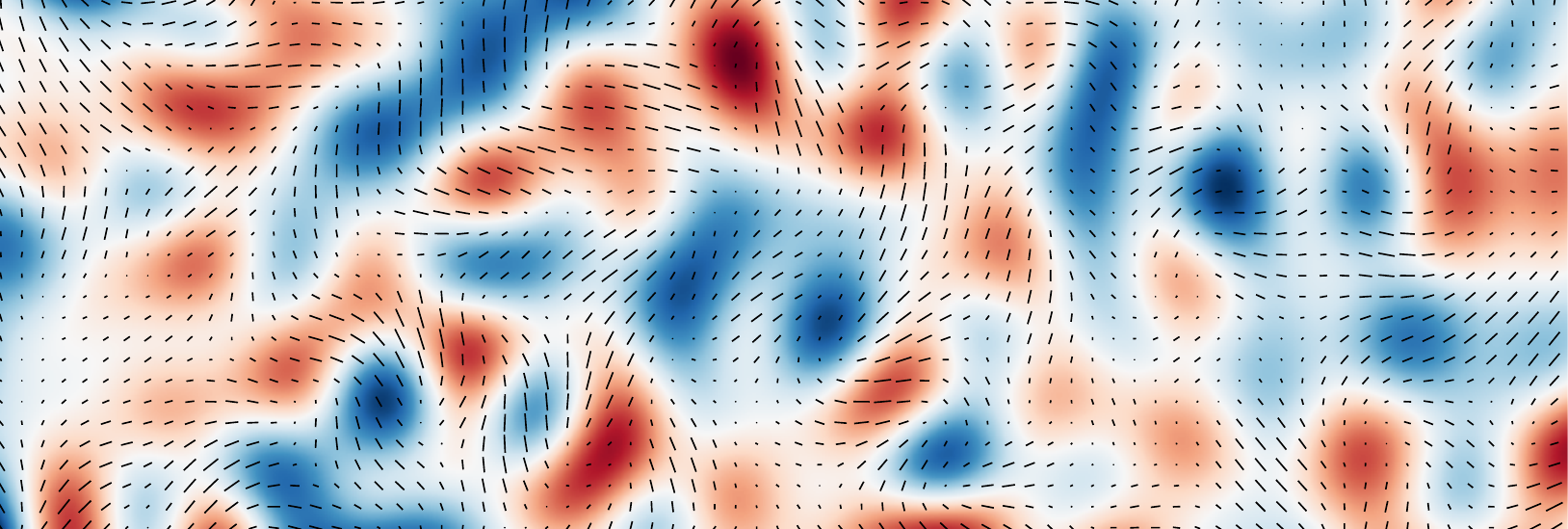}
} \\[1em] 
\subfloat[$\eta_{t}= 0.5\eta_{\text{max}}$\label{fig:BmodeOpen5}]{
    \includegraphics[width=0.5\textwidth]{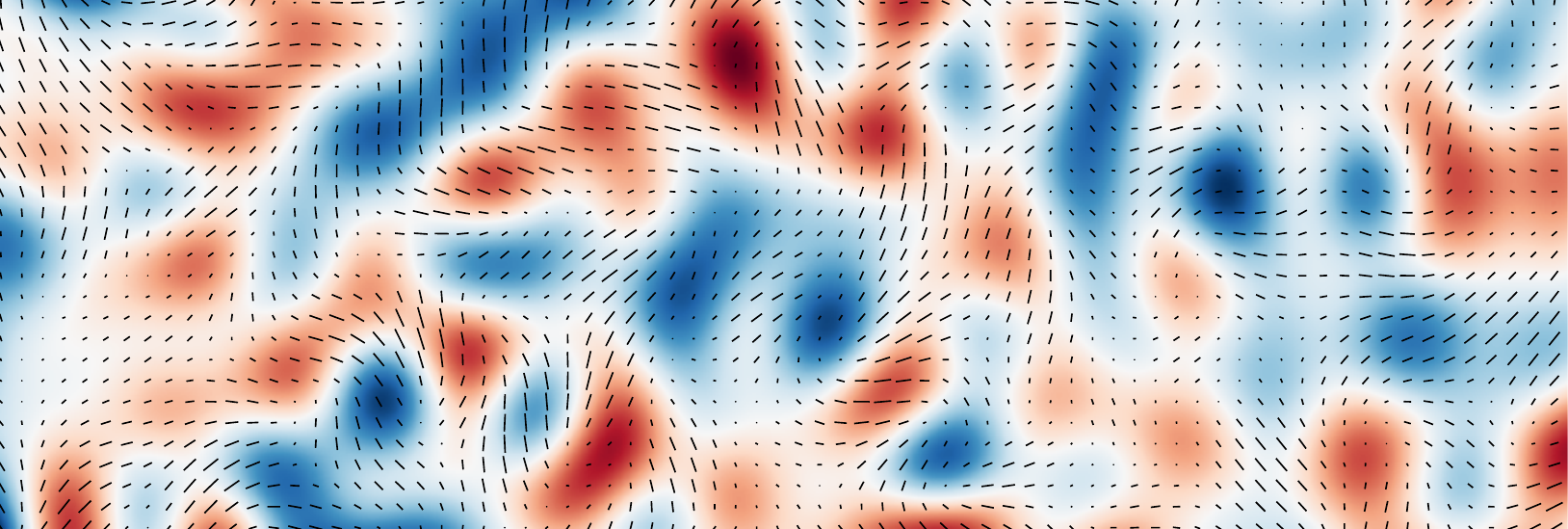}
}
\subfloat[$\eta_{t}=\eta_{\text{max}}$\label{fig:BmodeOpenMax}]{
    \includegraphics[width=0.5\textwidth]{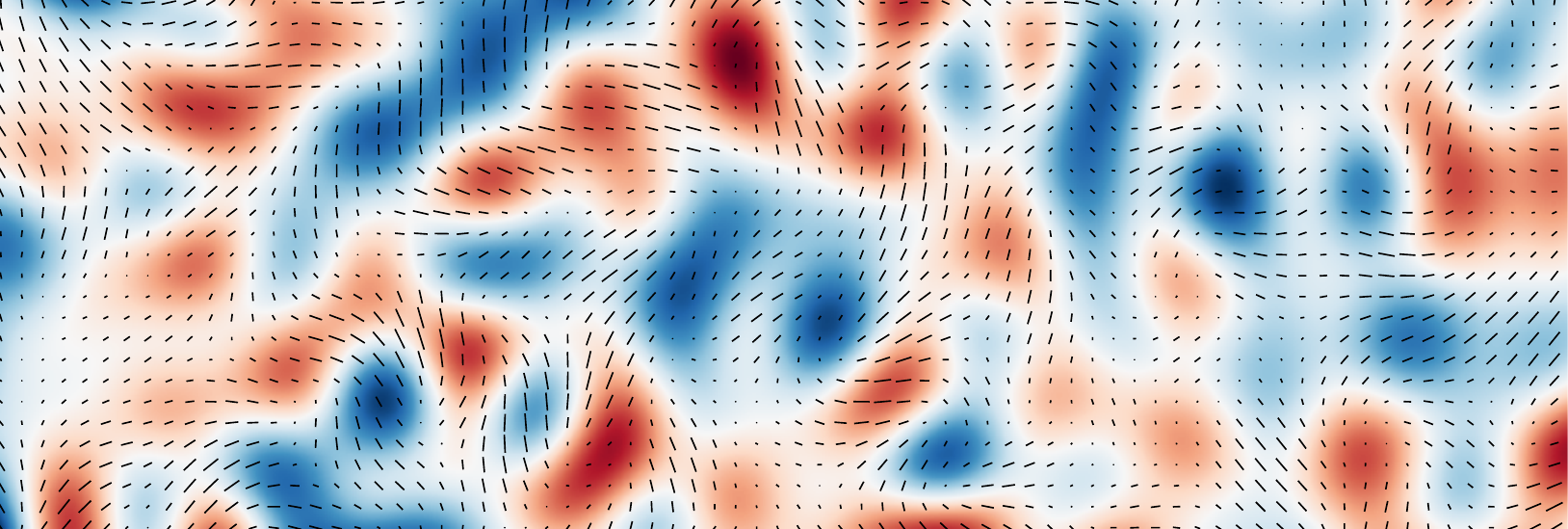}
} \\[1em] 
\subfloat[$K\Lambda$CDM\label{fig:BmodeKLCDM}]{
    \includegraphics[width=0.5\textwidth]{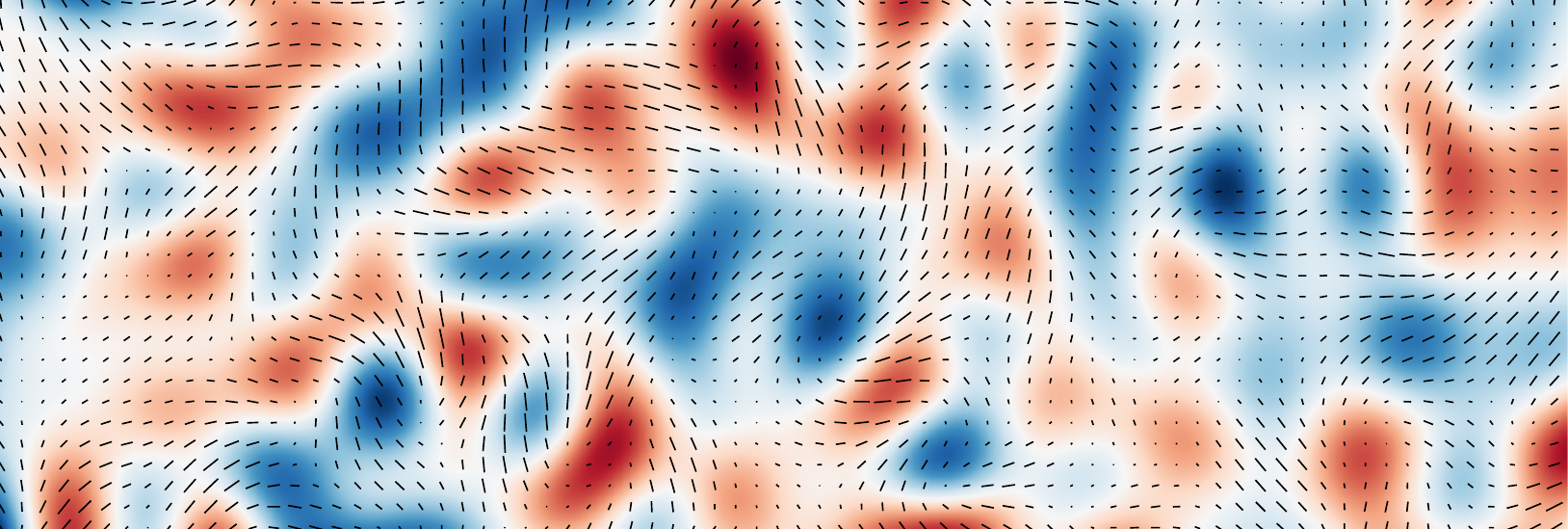}
}
\caption{\label{fig:OpenPolarisation} Simulated open universes $(K=-1)$ CMB $B$-mode polarisation maps of four chosen values of transition time $\eta_t \in \{(0.03,0.1,0.5,1)\times \eta_{\text{max}} \}$ where $\eta_{\text{max}} = \pi/4$, and one for fiducial open $K\Lambda$CDM. Each image is a $15^{\circ}\times 5^{\circ}$ flat-sky projection capturing the low multipole ($< 500$) signatures of the $B$-mode polarisation. The polarisation maps have the colourbar truncated to $\pm 7 \times 10^{-5} \mu K^2$.}
\end{figure*}

\begin{figure*}[h!]
\captionsetup[subfigure]{labelformat=empty}
\centering
\subfloat[$\mathcal{D}^{BB}_{\ell,K\Lambda CDM} -\mathcal{D}^{BB}_{\ell,0.03\eta_{\text{max}}}$\label{fig:ClosedDifferenceeta03}]{
    \includegraphics[width=0.4\textwidth]{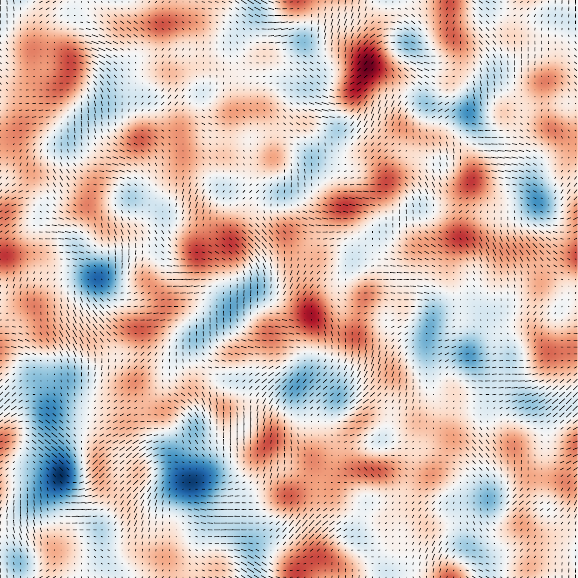}
}
\hspace{0.6cm}
\subfloat[$\mathcal{D}^{BB}_{\ell,K\Lambda CDM} -\mathcal{D}^{BB}_{\ell,0.1\eta_{\text{max}}}$\label{fig:ClosedDifferenceeta1}]{
    \includegraphics[width=0.4\textwidth]{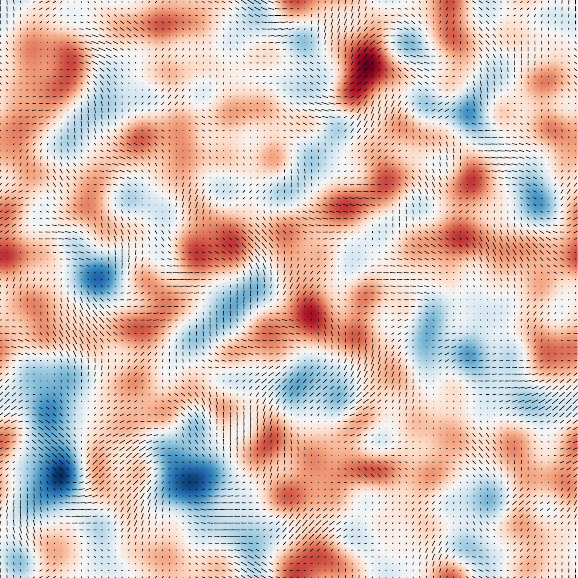}
} \\[1em] 
\subfloat[$\mathcal{D}^{BB}_{\ell,K\Lambda CDM} -\mathcal{D}^{BB}_{\ell,0.5\eta_{\text{max}}}$\label{fig:ClosedDifferenceeta5}]{
    \includegraphics[width=0.4\textwidth]{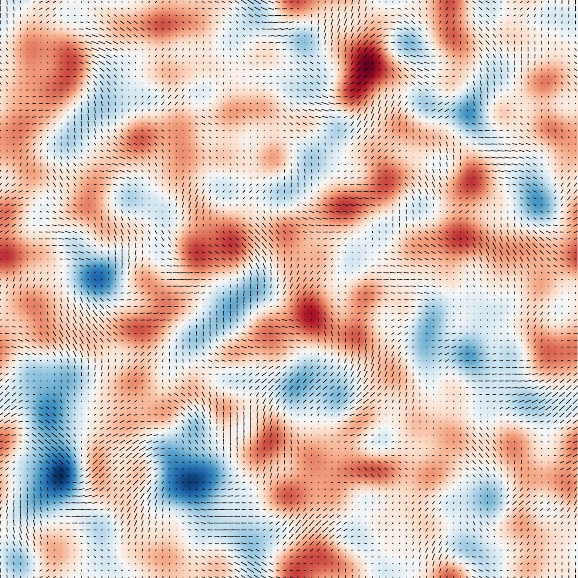}
}
\hspace{0.6cm}
\subfloat[$\mathcal{D}^{BB}_{\ell,K\Lambda CDM} -\mathcal{D}^{BB}_{\ell,\eta_{\text{max}}}$\label{fig:ClosedDifferenceetamax}]{
    \includegraphics[width=0.4\textwidth]{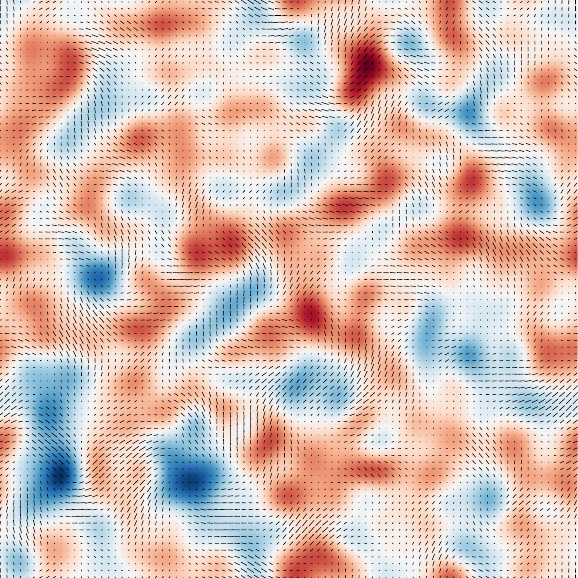}
}
\caption{Simulated closed universe $(K=+1)$ CMB $B$-mode polarisation map differences between closed $K\Lambda$CDM spectra and the analytic spectra for the given values of transition time $\eta_{t}$. We generate the polarisation maps taken from the analytical spectra $\mathcal{D}^{BB}_{\ell,K\Lambda CDM} -\mathcal{D}^{BB}_{\ell,\eta_{t}}$, where $\eta_t \in \{(0.03,0.1,0.5,1)\times \eta_{\text{max}} \}$ and $\eta_{\text{max}} = \pi/4$. Each image is a $17^{\circ}\times 17^{\circ}$ flat-sky projection to better capture the low multipole ($< 500$) signatures of the $B$-mode polarisation. The labels also identify the colourbar scale for each polarisation map. Each map has the colourbar truncated to order $\mathcal{O}(10^{-7}) \mu K^{2}$.}
\label{fig:ClosedPolarisationDifference}
\end{figure*}

\begin{figure*}[h!]
\captionsetup[subfigure]{labelformat=empty}
\centering
\subfloat[$\mathcal{D}^{BB}_{\ell,K\Lambda CDM} -\mathcal{D}^{BB}_{\ell,0.03\eta_{\text{max}}}$, $\mathcal{O}(10^{-9}) \mu K^2$\label{fig:OpendDifferenceeta03}]{
    \includegraphics[width=0.4\textwidth]{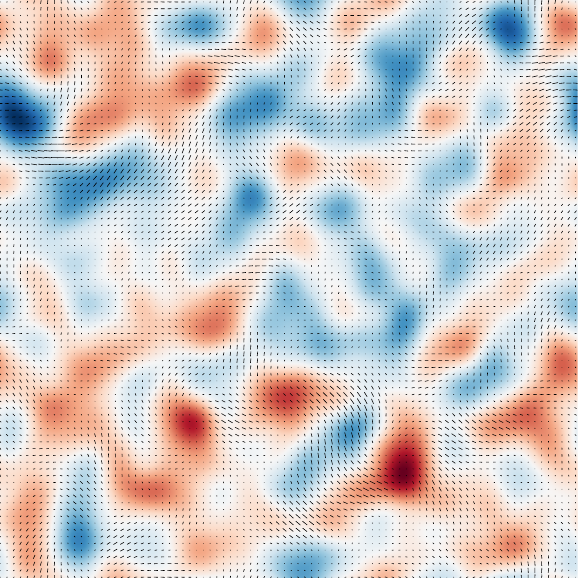}
}
\hspace{0.6cm}
\subfloat[$\mathcal{D}^{BB}_{\ell,K\Lambda CDM} -\mathcal{D}^{BB}_{\ell,0.1\eta_{\text{max}}}$, $\mathcal{O}(10^{-8}) \mu K^2$\label{fig:OpenDifferenceeta1}]{
    \includegraphics[width=0.4\textwidth]{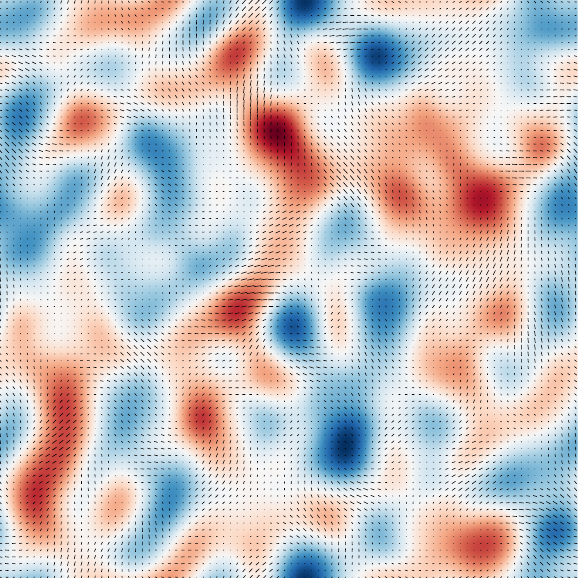}
} \\[1em] 
\subfloat[$\mathcal{D}^{BB}_{\ell,K\Lambda CDM} -\mathcal{D}^{BB}_{\ell,0.5\eta_{\text{max}}}$, $\mathcal{O}(10^{-10}) \mu K^2$\label{fig:OpenDifferenceeta5}]{
    \includegraphics[width=0.4\textwidth]{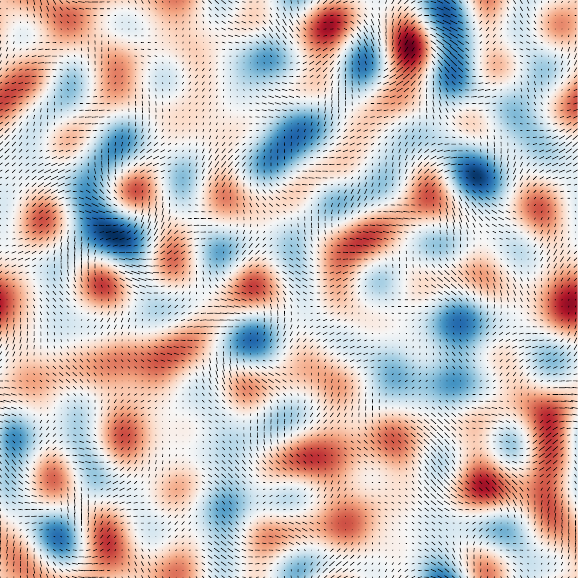}
}
\hspace{0.6cm}
\subfloat[$\mathcal{D}^{BB}_{\ell,K\Lambda CDM} -\mathcal{D}^{BB}_{\ell,\eta_{\text{max}}}$, $\mathcal{O}(10^{-11}) \mu K^2$\label{fig:OpenDifferenceetamax}]{
    \includegraphics[width=0.4\textwidth]{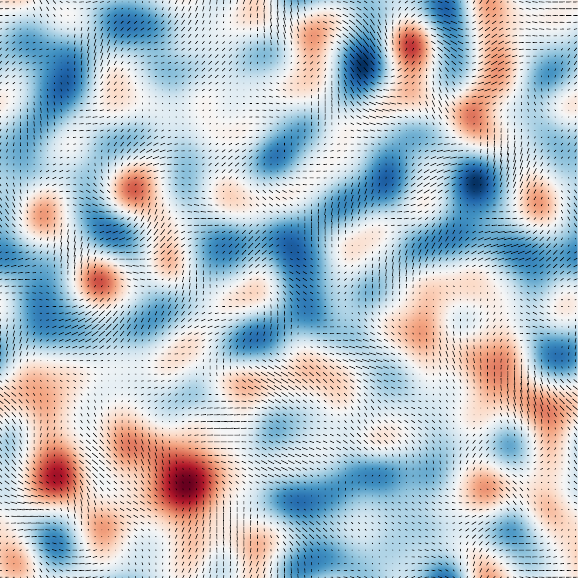}
}
\caption{Simulated open universe $(K=-1)$ CMB $B$-mode polarisation map differences between open $K\Lambda$CDM spectra and the analytic spectra for the given values of transition time $\eta_{t}$. We generate the polarisation maps taken from the analytical spectra $\mathcal{D}^{BB}_{\ell,K\Lambda CDM} -\mathcal{D}^{BB}_{\ell,\eta_{t}}$, where $\eta_t \in \{(0.03,0.1,0.5,1)\times \eta_{\text{max}} \}$ and $\eta_{\text{max}} = \pi/4$. Each image is a $17^{\circ}\times 17^{\circ}$ flat-sky projection to better capture the low multipole ($< 500$) signatures of the $B$-mode polarisation. The labels also identify the colourbar scale for each polarisation map.}
\label{fig:OpenPolarisationDifference}
\end{figure*}

\clearpage
\bibliographystyle{unsrtnat}
\bibliography{references}

\end{document}